\PassOptionsToPackage{table,dvipsnames}{xcolor}
\documentclass[sigconf]{acmart}
\usepackage{xcolor}

\usepackage{array,makecell,multirow}
\usepackage{float}                
\usepackage{acmart-taps}       

\copyrightyear{2026}
\acmYear{2026}
\setcopyright{cc}
\setcctype{by}
\acmConference[CHI '26]{Proceedings of the 2026 CHI Conference on Human Factors in Computing Systems}{April 13--17, 2026}{Barcelona, Spain}
\acmBooktitle{Proceedings of the 2026 CHI Conference on Human Factors in Computing Systems (CHI '26), April 13--17, 2026, Barcelona, Spain}
\acmPrice{}
\acmDOI{10.1145/3772318.3791410}
\acmISBN{979-8-4007-2278-3/2026/04}

\definecolor{ContestantBg}{HTML}{CCE8FF}
\definecolor{ProblemSetterBg}{HTML}{FFE8CC}
\definecolor{CoachBg}{HTML}{E6FFCC}
\definecolor{PlatformBg}{HTML}{F0CCFF}
\definecolor{myblue}{HTML}{CCE5FF}
\definecolor{Lblue}{HTML}{E6F3FF}

\makeatletter
\def\C#1{\allowbreak\hyperlink{C#1}{\colorbox{ContestantBg}{\strut C#1}}}
\def\Cc#1{\allowbreak\hyperlink{C#1}{\colorbox{ContestantBg}{\strut \textit{C#1}}}}
\newcommand{\Pp}[1]{\allowbreak\hyperlink{P#1}{\colorbox{ProblemSetterBg}{\strut P#1}}}
\def\Ppp#1{\allowbreak\hyperlink{P#1}{\colorbox{ProblemSetterBg}{\strut \textit{P#1}}}}
\def\T#1{\allowbreak\hyperlink{T#1}{\colorbox{CoachBg}{\strut T#1}}}\def\Tt#1{\allowbreak\hyperlink{T#1}{\colorbox{CoachBg}{\strut \textit{T#1}}}}
\newcommand{\G}[1]{\allowbreak\hyperlink{G#1}{\colorbox{PlatformBg}{\strut G#1}}}
\def\Gg#1{\allowbreak\hyperlink{G#1}{\colorbox{PlatformBg}{\strut \textit{G#1}}}}

\newcommand{\Cid}[1]{\hypertarget{C#1}{\colorbox{ContestantBg}{\strut C#1}}}
\newcommand{\Pid}[1]{\hypertarget{P#1}{\colorbox{ProblemSetterBg}{\strut P#1}}}
\newcommand{\Tid}[1]{\hypertarget{T#1}{\colorbox{CoachBg}{\strut T#1}}}
\newcommand{\Gid}[1]{\hypertarget{G#1}{\colorbox{PlatformBg}{\strut G#1}}}

\makeatother

\definecolor{newGreen}{HTML}{009900}
\definecolor{newCyan}{HTML}{2A9BC9}
\definecolor{newOrange}{HTML}{E57300}

\newcommand{\handleGray}[1]{\textcolor{gray}{#1}}                 
\newcommand{\handleGreen}[1]{\textcolor{newGreen}{#1}}      
\newcommand{\handleCyan}[1]{\textcolor{newCyan}{#1}}        
\newcommand{\handleBlue}[1]{\textcolor{blue}{#1}}                 
\newcommand{\handlePurple}[1]{\textcolor{violet}{#1}}             
\newcommand{\handleOrange}[1]{\textcolor{newOrange}{#1}}    
\newcommand{\handleRed}[1]{\textcolor{red}{#1}}                   
\newcommand{\handleLG}{\textcolor{black}{L}\textcolor{red}{egendary Grandmaster}}

\aptLtoXcmd{}{
\captionsetup{font=small, skip=4pt}
\setlength{\textfloatsep}{8pt plus 2pt minus 2pt}
\setlength{\floatsep}{8pt plus 2pt minus 2pt}
\setlength{\intextsep}{8pt plus 2pt minus 2pt}}

\makeatletter
\@ifundefined{color@ProblemSetterBg}{\definecolor{ProblemSetterBg}{HTML}{FFE8CC}}{}
\@ifundefined{color@CoachBg}{\definecolor{CoachBg}{HTML}{E6FFCC}}{}
\@ifundefined{color@PlatformBg}{\definecolor{PlatformBg}{HTML}{F0CCFF}}{}
\makeatother

\aptLtoXcmd{}{
\renewcommand{\arraystretch}{1.15}
\setlength{\tabcolsep}{4pt}
\setlength{\arrayrulewidth}{0.5pt}
\setlength{\textfloatsep}{6pt plus 2pt minus 2pt}
\setlength{\floatsep}{6pt plus 2pt minus 2pt}
\setlength{\intextsep}{6pt plus 2pt minus 2pt}}

\small
\aptLtoXcmd{}{
\renewcommand{\arraystretch}{1.12}
\setlength{\tabcolsep}{3.5pt}}



\begin{document}
\title[FAIR: Framing AI’s Role in Programming Competitions]{FAIR: Framing AI’s Role in Programming Competitions --- Understanding How LLMs Are Changing the Game in Competitive Programming}

\author{Dongyijie Primo Pan}
\email{dpan750@connect.hkust-gz.edu.cn}
\affiliation{\institution{Hong Kong University of Science and Technology (Guangzhou)} \city{Guangzhou} \state{Guangdong} \country{China}}

\author{Lan Luo}
\authornotemark[1]
\email{lluo476@connect.hkust-gz.edu.cn}
\affiliation{\institution{Hong Kong University of Science and Technology (Guangzhou)} \city{Guangzhou} \state{Guangdong} \country{China}}

\author{Ji Zhu}
\authornote{These authors contributed equally to this work}
\email{zhuji@cuc.edu.cn}
\affiliation{
  \institution{Communication University of China}
  \state{Beijing}
  \country{China}
}

\author{Zhiqi Gao}
\email{zhiqigao@link.cuhk.edu.cn}
\affiliation{\institution{The Chinese University of Hong Kong, Shenzhen} \city{Shenzhen} \state{Guangdong} \country{China}}

\author{Xin Tong}
\authornote{Corresponding author}
\email{xtong@hkust-gz.edu.cn}
\affiliation{\institution{Hong Kong University of Science and Technology (Guangzhou)} \city{Guangzhou} \state{Guangdong} \country{China}}

\author{Pan Hui}
\authornotemark[2]
\email{panhui@hkust-gz.edu.cn}

\affiliation{\institution{Hong Kong University of Science and Technology (Guangzhou)} \city{Guangzhou} \state{Guangdong} \country{China}}
\affiliation{\institution{Hong Kong University of Science and Technology} \country{Hong Kong SAR}}

\renewcommand{\shortauthors}{Pan et al.}
\begin{abstract}
This paper investigates how large language models (LLMs) are reshaping competitive programming. The field functions as an intellectual contest within computer science education and is marked by rapid iteration, real-time feedback, transparent solutions, and strict integrity norms. Prior work has evaluated LLMs performance on contest problems, but little is known about how human stakeholders—contestants, problem setters, coaches, and platform stewards—are adapting their workflows and contest norms under LLMs-induced shifts. At the same time, rising AI-assisted misuse and inconsistent governance expose urgent gaps in sustaining fairness and credibility. Drawing on 37 interviews spanning all four roles and a global survey of 207 contestants, as well as an API-based crawl of Codeforces contest logs (2022–2025) for quantitative analysis, we contribute: (i) an empirical account of evolving workflows, (ii) an analysis of contested fairness norms, and (iii) a chess-inspired governance approach with actionable measures—real-time LLMs checks in online contests, peer co-monitoring and reporting, and cross-validation against offline performance—to curb LLMs-assisted misuse while preserving fairness, transparency, and credibility.
\end{abstract}



\maketitle
\section{Introduction}

Every weekend, tens of thousands of users worldwide log into online programming platforms (e.g., \textbf{Codeforces}) to compete in timed contests. Under strict time and memory limits with instant online-judge feedback, contestants race to craft correct algorithms in a highly interactive human–computer, high-stakes setting. These platforms host millions and run regular rounds that mirror onsite rigor, anchoring computing education and curricula while cultivating algorithmic thinking, resilience, and collaboration~\cite{yuen2023competitive,mirzayanovCodeforcesEducationalPlatform2020}. Flagship events like the \textbf{International Collegiate Programming Contest (ICPC)} and the \textbf{International Olympiad in Informatics (IOI)} sustain a global community and shared norms~\cite{halim2013competitive,10.1145/2839509.2844632,opmanis2006some}. In this ecosystem, competitive programming serves both pedagogy and a talent pipeline: academia and top tech firms prize outstanding performers, with medalists actively recruited and companies hosting their own contests (e.g., \textbf{Google Code Jam, Meta Hacker Cup})~\cite{mirzayanovCodeforcesEducationalPlatform2020,dymchenko2015declaratively}.

Against this backdrop, LLMs are changing the rules of competitive programming—a field long premised on contestants solving problems without outside assistance—by compressing iteration cycles via code synthesis, bug localization, and targeted test generation ~\cite{mozannar2024realhumaneval}. Concurrently, across industry and academia, competitive-programming outcomes are increasingly used as capability indices for LLMs' code-generation systems~\cite{li2022competition,openai2025competitiveprogramminglargereasoning,Jain2024LiveCodeBench:,souza2025codegenerationsmalllanguage,10.5555/3737916.3739323,dumitran2024evaluating,zheng2025livecodebenchproolympiadmedalists}, with \textbf{Codeforces} rating serving as the de facto gold standard~\cite{huang-etal-2024-competition}. As model capabilities rise, high-profile scandals over AI-assisted submissions—including at national-level contests—have surfaced~\cite{CEMC_CCC_2025,codeforces2023cheating,zaobao2025cheating}, and the community faces unsettled boundaries: contestants wrestle with when and how AI assistance is permissible; problem setters debate whether to adopt anti-AI task strategies; and coaches and platform stewards must detect and adjudicate AI-enabled gains without uniform standards~\cite{10.1145/3706598.3713357,10.1145/3674805.3686689}. Despite the emphasis on ratings, there is still limited empirical work on how stakeholders actually respond on the ground—how workflows change and how fairness and governance decisions get made~\cite{dumitran2024evaluating,huang-etal-2024-competition,watanobe2022online}.

Prior HCI and CSE research has examined AI use in online education and proctored exams, showing that generative models can support personalized feedback, automated grading, and scaffolding for novice programmers, but also raise risks of over-reliance, diminished critical thinking, and integrity violations~\cite{10.1145/3544548.3580919,alfageeh2025prompts}. Detection and governance studies likewise highlight how plagiarism and AI-assisted cheating challenge trust in digital learning environments, yet technical safeguards are easily evaded and rules remain inconsistent across platforms~\cite{ullah2021intelligent}. What remains underexplored is how these dynamics unfold in competitive programming, a domain that differs markedly from typical educational settings: contestants receive instant in-contest feedback; all solutions are published afterward, encouraging peer learning; and integrity norms are actively enforced by the community~\cite{halim2013competitive}. Unlike coursework or exams, which rarely offer comparable interactivity~\cite{10.1145/3706598.3713357,10.1145/3544548.3580919}, contests operate as highly interactive, high-stakes arenas. Moreover, while classroom-based programming education relies on teachers to scaffold progression across levels, advanced contestants in competitive programming typically progress through long-term self-study and community peer practice once they surpass the novice stage\cite{liu2008training,park2007roles}. The rapid spread of AI-generated solutions therefore raises urgent and distinctive challenges for safeguarding contest integrity—challenges that existing HCI and CSE literature has yet to address.

To address these gaps, we pose three research questions:
\begin{itemize}
\item \textbf{RQ1—Workflow}: How are LLMs reshaping the day-to-day workflows of contestants, problem setters, coaches and platform stewards?
\item \textbf{RQ2—Fairness}: How do stakeholders define and contest the boundary between acceptable AI assistance and cheating?
\item \textbf{RQ3—Governance}: How do online platforms and their communities co-create and iterate AI usage rules to safeguard credibility?
\end{itemize}

\noindent To answer these questions, we conducted an empirical study that combines \textbf{37 in-depth interviews} across four roles—contestants (from \textbf{world champions} to \textbf{novices}), problem setters, coaches, and platform stewards—with a \textbf{global survey of 207 contestants} from diverse countries and regions. By drawing on both interviews and survey perspectives, we provide a comprehensive account of how LLMs are transforming competitive programming and what governance responses may be needed.  

\noindent This paper makes the following contributions:
\begin{itemize}
    \item \textbf{Empirical insights into evolving workflows}: From 37 interviews and a global survey of 207 contestants, we document how LLMs are reshaping training, contest participation, problem setting, and platform stewardship.  
    \item \textbf{Cross-role perspectives on fairness}: We analyze how different stakeholders—contestants, problem setters, coaches, and platform stewards—define and contest the boundary between legitimate assistance and cheating.  
    \item \textbf{Chess-inspired governance approach}: Synthesizing these findings, we propose a governance approach for programming contests, inspired by how mind sports regulate AI use, offering actionable guidelines to safeguard fairness and credibility in the AI era.  
\end{itemize}

\section{Background and Related Work}

\subsection{Background on Competitive Programming Rules}
\label{subsec:contest-rules}

\subsubsection{Contest formats (ICPC, IOI, Codeforces).}

Competitive programming contests come in different formats. The \textbf{International Collegiate Programming Contest (ICPC)} is a team-based competition for university students, where teams of three solve 8--12 problems in 5 hours using a single computer. Scoring is \emph{all-or-nothing}, with a 20-minute penalty for each incorrect submission~\cite{10.1145/2839509.2844632}. In contrast, the \textbf{International Olympiad in Informatics (IOI)} is an individual contest for high school students, held over two days with 3 problems each day. Partial credit (up to 100 points per problem) is awarded, with no penalties for incorrect submissions~\cite{opmanis2006some}.

To simulate the real competition environment for daily training, numerous online programming platforms have emerged. Some of the most widely used platforms include \textit{Codeforces} (Russia), \textit{Luogu} (China), and \textit{AtCoder} (Japan). Among them, \textit{Codeforces} stands out as the largest and most influential platform~\cite{mirzayanovCodeforcesEducationalPlatform2020}, hosting individual rounds ($\approx$2 hours, $\approx$5 problems) with ICPC-style scoring (full-solution required; small penalties for wrong submissions). Because these contests are remote and asynchronous across time zones, they pose distinct governance and fairness challenges compared to proctored, in-person contests.

Unlike online platforms, major onsite contests such as ICPC and IOI are conducted under strict proctoring conditions: participants are searched before entry and work in isolated local networks without Internet access. These safeguards virtually eliminate the possibility of external assistance, including AI tools. By contrast, online contests are hosted in uncontrolled home or campus environments without such physical restrictions, making questions of AI use and fairness particularly salient for our study.

\subsubsection{Online judging and rating systems.}

A defining feature of competitive programming—spanning both onsite contests such as ICPC and IOI and online platforms such as Codeforces—is their reliance on an \textbf{online judge (OJ) system}. Contestants submit code to the OJ, which immediately compiles and runs it against hidden test cases. This real-time feedback fosters an iterative “submit–debug–resubmit” workflow that demands both technical skill and mental resilience~\cite{yuen2023competitive}. Submission outcomes typically include \emph{Accepted} (correct output within limits), \emph{Wrong Answer} (failed test cases), or \emph{Time Limit Exceeded} (runtime overrun), each shaping scoring and pacing. Beyond individual verdicts, OJ systems also provide a \textbf{real-time public ranking board} that continuously updates contestants’ standings, further influencing their problem-solving order and strategic decisions during contests.

Beyond individual submissions, online platforms maintain rating systems that track long-term performance. Among them, \textit{Codeforces} uses an Elo-style rating updated after each round~\cite{pelanek2016applications}, mapping users into colored handle tiers (Table~\ref{tab:cf-rating}). These tiers concisely indicate skill and are widely referenced in both community and research contexts; they also serve as a benchmark axis when evaluating LLMs on programming tasks~\cite{huang-etal-2024-competition}.

\begin{table}[htbp]
  \centering
  \caption{Codeforces user tiers and rating intervals. Handle colors correspond to the official Codeforces tier badges.}
  \small
  \setlength{\tabcolsep}{6pt}
  \renewcommand{\arraystretch}{1.1}
  \begin{tabular}{|l|c|}
    \hline
    \textbf{Tier (colored handle)} & \textbf{Rating Range} \\
    \hline
    \handleGray{Newbie} & $\leq 1199$ \\ \hline
    \handleGreen{Pupil} & 1200--1399 \\ \hline
    \handleCyan{Specialist} & 1400--1599 \\ \hline
    \handleBlue{Expert} & 1600--1899 \\ \hline
    \handlePurple{Candidate Master} & 1900--2099 \\ \hline
    \handleOrange{Master} & 2100--2299 \\ \hline
    \handleOrange{International Master} & 2300--2399 \\ \hline
    \handleRed{Grandmaster} & 2400--2599 \\ \hline
    \handleRed{International Grandmaster} & 2600--2999 \\ \hline
    \handleLG & $\geq 3000$ \\ \hline
  \end{tabular}
  \Description{Table listing Codeforces rating tiers with corresponding ranges. 
  Newbie: 1199 or below; Pupil: 1200--1399; Specialist: 1400--1599; Expert: 1600--1899; Candidate Master: 1900--2099; 
  Master: 2100--2299; International Master: 2300--2399; Grandmaster: 2400--2599; 
  International Grandmaster: 2600--2999; Legendary Grandmaster: 3000 or higher. 
  Colors shown in the first column match Codeforces official handle colors.}
  \label{tab:cf-rating}
\end{table}

\subsection{Related Work}
\subsubsection{LLMs Performance on Competitive Programming Benchmarks}

In recent years, programming contest scores have become a widely used benchmark in both industry and academia for evaluating the coding capabilities of LLMs~\cite{huang-etal-2024-competition}. However, these headline results often conceal significant methodological issues~\cite{li2022competition,openai2025competitiveprogramminglargereasoning,Jain2024LiveCodeBench:,10.5555/3737916.3739323}. Some evaluations only report \textbf{pass@1}\footnote{\textbf{pass@}$n$ is the proportion of problems for which at least one of $n$ independently generated solutions passes the problem; \textbf{pass@1} uses only the model's first attempt.}, which ignores the interactive nature of real contests, where contestants iteratively refine their solutions in response to feedback. Others use \textbf{pass@k}~\cite{wang2024performance}, where increasing \textbf{k} greatly inflates success rates through multiple attempts~\cite{10.1007/s11704-024-40415-9}; in actual contests, penalties for incorrect submissions discourage such guessing and force careful decision-making~\cite{10.1145/2839509.2844632}. Real-time ranking updates—missing from most LLMs evaluations—also shape contestants’ problem-solving order and strategies~\cite{10.1145/3233390}. More broadly, current benchmarks cannot fully capture how human contestants adapt their \textbf{workflows} under pressure or adjust strategies in response to feedback, meaning that evaluations systematically overlook the \textbf{human factors} that define competitive programming.

As a result, many question the validity of “model ratings,” which lack standardized, realistic foundations. Scores can vary widely with prompt design, test-set selection, and are susceptible to prompt engineering or training-set leakage~\cite{dumitran2024evaluating}. For example, Huang et~al.\ found GPT-4 performed significantly worse on Codeforces problems released after 2022, suggesting possible memorization of earlier data~\cite{huang-etal-2024-competition}. The absence of standardized, publicly agreed-upon evaluation protocols allows performance to be reported under heterogeneous and potentially favorable conditions, making headline scores prone to inflation in promotional materials. 

To address these issues, Zheng et~al.\ introduced \textbf{LiveCodeBench Pro}, a benchmark labeled by IOI and ICPC medalists to assess models on unseen, high-difficulty problems~\cite{zheng2025livecodebenchproolympiadmedalists}. In August 2025, OpenAI claimed that \textbf{GPT-5-high} achieved a simulated \textbf{6th-place ranking among human contestants at IOI 2025}~\cite{OpenAI2025IOI}, yet only attained a \textbf{5.9\%} success rate on LiveCodeBench Pro’s high-difficulty problems. While this represents measurable progress, it remains far below top human performance. As the leader \textbf{LiveCodeBench Pro}, Saining Xie noted on social media, “The AlphaGo moment for AI programming has not yet arrived.”~\cite{Xie2025AlphaGo}

Taken together, these findings highlight a persistent gap between leaderboard metrics and genuine competitive programming performance.

\subsubsection{The Impact of LLMs on Programming Workflows}

Although there is growing excitement around higher Codeforces ratings achieved by newly released models from tech companies, these ratings rarely translate directly into tangible benefits for real-world software development tasks~\cite{mozannar2024realhumaneval, wen2024language}. Nevertheless, the research landscape reveals that LLMs are reshaping programming workflows by automating routine tasks, supporting collaborative coding, and democratizing access to software development~\cite{jiang2022discovering,10.1145/3613904.3641936,Chen2024Learning,Ross2023The}. In-IDE AI plugins~\cite{Sergeyuk2024In-IDE}, proactive AI assistants~\cite{10.1145/3706598.3714002}, and interaction methods that more closely resemble real-world thinking processes~\cite{10.1145/3586183.3606719,10.1145/3672539.3686324} open up new possibilities for enhancing both the productivity and creative thinking of programmers. \textbf{Microsoft} team's experiment shows that developers using Copilot completed industrial programming tasks 55.8\% faster than those who did not use the tool~\cite{peng2023impactaideveloperproductivity}. However, Fakhoury et al.'s work shows the effectiveness is highly dependent on prompt design, user expertise, and the complexity of programming tasks~\cite{Fakhoury2024LLM-Based}. While LLMs democratize access to programming and lower barriers for non-experts, they also raise concerns about code quality, security, and the need for human oversight~\cite{Liu2023No,Gao2024The,Hou2023Large,Zhong2023Can}.

Interestingly, while extensive research has been conducted on the impact of large language models (LLMs) in industrial and data analysis programming, there is a notable lack of studies examining how LLMs influence the workflows of competitive programming participants and problem setters. This gap underscores the limitations of existing research, particularly in the realm of competitive programming, which serves as a benchmark for evaluating the "programming" capabilities of large models. Therefore, our \textbf{RQ1} aims to address this void by delving into the applications and impacts of LLMs in competitive programming.

\subsubsection{LLMs and Programming Education}

Most empirical HCI and CS Education studies still probe LLMs in \emph{early-stage} programming practice~\cite{10.1145/3706598.3713357,10.1145/3706598.3714154,10.1145/3568813.3600139}. Prior work shows LLMs can act as pedagogical agents, teachable tutees, feedback generators, and tools for automating grading and content creation ~\cite{10.1145/3613904.3642349,10.1145/3687038,10.1145/3613904.3642773,10.1145/3587102.3588785}, offering scalable, personalized feedback to improve engagement and support code generation, debugging, and curriculum design ~\cite{10.1145/3545945.3569823,10.1145/3576123.3576134,10.1145/3649217.3653533,10.1145/3617367,10.1145/3526113.3545659}. Yet these benefits face challenges—over-reliance, diminished critical thinking, academic integrity issues, and limits in nuanced, context-aware support ~\cite{10.1109/TVCG.2024.3456363,10.1145/3626252.3630958,10.1145/3491101.3519665,10.1145/3568813.3600138}. Existing research largely focuses on novice instruction and industry-oriented tasks~\cite{10.1145/3544548.3580919,Song2024A}, leaving a gap in understanding LLMs' effectiveness in advanced programming education, such as competitive programming with strict time limits, high complexity, and algorithmic depth~\cite{10.1145/3587102.3588773,wolfer2023qualitative,10.1145/3587102.3588792}. Unlike classroom-based computer science education, where instructors scaffold progression across levels, advanced contestants in programming contests typically rely on long-term self-study, community resources, and peer practice to continue improving once they surpass the novice stage~\cite{liu2008training,park2007roles}. Studies also rarely examine effects on educators’ workflows—lesson preparation, problem-setting, and nuanced assessment~\cite{Macneil2022Automatically,Lejmbach2024Using,hoq2025facilitatinginstructorsllmcollaborationproblem}—while cheating, academic integrity concerns, and unclear usage boundaries~\cite{10.1145/3544548.3580919,Song2024A} persist in competitive programming. Our \textbf{RQ1} and \textbf{RQ2} address these issues.

\subsubsection{
    Governance and Fairness in Online Education and Community Platforms}

The governance and fairness of online education and community platforms have become pressing concerns as digital learning proliferates. While such platforms broaden access, they may also exacerbate inequalities without proper oversight~\cite{Tate2022Equity,Jhaver2021Decentralizing}. Examinations and competitions are particularly vulnerable: weak governance heightens cheating risks~\cite{Noorbehbahani2022A,Newton2023How}. In programming education, common issues include plagiarism and AI-assisted solutions~\cite{10.1145/3210713.3210724}. Although plagiarism and AI-detection algorithms exist~\cite{Novak2019Source-code,7816515,10.1145/3657604.3662046,salim-etal-2024-impeding,11029804}, studies show they are easily evaded~\cite{gritsai2025are,dickey2025failure}. Some propose controlled use of LLMs in contests~\cite{amsen2024cfai}, yet concerns over fairness, access disparities, and model quality fuel broad resistance~\cite{Kwak2024Bridging,yu2024whose}. Consequently, most major platforms continue to ban generative AI during contests~\cite{Mirzayanov2024CFPolicy}.

Thanks to the open and community-governed nature of competitive programming platforms, cheaters who evade system detection are often reported by users and have their results invalidated. Some communities even maintain third-party reputation systems\footnote{For example, \url{https://cfcheatdetector.netlify.app/}
 and \url{https://cf-cheater-database.vercel.app/}}~\cite{resnick2000reputation}
. Such practices highlight that governance in these platforms differs from conventional laws, instead reflecting community norms and rules~\cite{10.1145/3613904.3642012,10.1145/1357054.1357227,fiesler2018reddit}. Prior HCI work further stresses the complexity of user-driven flagging: empowering members with transparent, customizable tools can foster collaborative moderation and legitimacy~\cite{10.1145/3379337.3415858}, while modular frameworks enable flexible, responsive governance~\cite{DeFilippi2020Modular}. Yet, the absence of unified AI policies across major platforms continues to fuel conflicts between cheating accusations and defenses, underscoring the urgent need for credible governance. Beyond competitive and educational settings, creative online communities—such as book fandoms~\cite{Li2024Fandom}, fan-fiction writers~\cite{Formosa2024Can}, and digital-art communities~\cite{Messer2024Co-creating} responding to image-generation models—are also embroiled in tensions around generative AI, but with fundamentally different concerns: debates center on authorship, attribution, and creative labor rather than cheating or rating inflation. Because our data derive from rule-bound contests where AI assistance is interpreted as violating explicit fairness norms, we treat these creative communities as adjacent yet conceptually distinct domains and do not generalize our governance proposals to co-authoring or artistic production settings. Our study therefore addresses \textbf{RQ2} and \textbf{RQ3} to synthesize consensus-driven perspectives in the community.

\section{Methods}

\noindent To address our three research questions---\textbf{RQ1 (workflows)}, \textbf{RQ2 (fairness)}, and \textbf{RQ3 (governance)}---we adopted a qualitative design, supported by survey triangulation. Our study consisted of two components: (i) \textbf{37 semi-structured interviews} with key stakeholders (contestants, coaches, problem setters, and platform stewards), which formed the primary dataset; and (ii) a \textbf{global online survey} of 207 active contestants, used to contextualize and extend interview findings with broader perspectives. Both instruments included open-ended questions, enabling integration of qualitative themes across datasets. 
All procedures received approval from our institutional review board (IRB). All interview transcripts and survey responses were stored on encrypted institutional servers with access restricted to the research team. 
Identifiers were removed during transcription, pseudonyms were assigned (e.g., \C{1}–\C{21}), and all data will be securely deleted five years after study completion in accordance with institutional policy.

\subsection{Semi-Structured Interviews}

\subsubsection{Participants.}
We conducted interviews with \textbf{37 unique individuals} across four roles: contestants, problem setters, coaches, and platform stewards. Among the \textbf{21 contestants}, only \textbf{4 were women} ($\approx$19\%), reflecting the well-documented gender imbalance in competitive programming~\cite{Yamaguchi2024The,Steegh2019Gender}. Contestants spanned the full spectrum of experience, from \textbf{novices} to \textbf{ICPC/IOI world champions}, and were drawn from major global regions. Beyond contestants, the sample included \textbf{7 problem setters}, \textbf{7 coaches}, and \textbf{8 platform stewards}; some held multiple roles. 

We report \textbf{unique individuals} (\textit{N}=37) and apply \textbf{role-based coding} in analysis. Detailed information on contestants is provided in Table~\ref{tab:contestants}, while profiles of problem setters, coaches, and platform stewards are listed in Appendix Tables~\ref{tab:A1-problemsetters}, \ref{tab:A2-coaches}, and \ref{tab:A3-platform}.

\subsubsection{Recruitment and Procedure.}
Participants were recruited purposively (May--Aug 2025) via calls on the \textit{Codeforces} forum, \textit{X}, and \textit{Reddit}, and further expanded through \textbf{snowball sampling} (participants recommending peers across roles and regions). During screening, we excluded individuals who (i) had not participated in any contests since the advent of LLMs, (ii) had contest experience too limited to demonstrate a basic understanding of contest rules and workflows, or (iii) coaches serving mainly in administrative or organizational roles without direct engagement in contest preparation or rules.

After screening, eligible participants joined either \textbf{video interviews} (Zoom) or completed a \textbf{written long-form questionnaire} when synchronous interviews were not feasible. Questionnaire respondents were typically those facing constraints such as large time-zone differences, limited English proficiency, or restricted access to stable video-conferencing tools. To ensure comparable richness, we incorporated \textbf{follow-up prompts and clarifying exchanges} via email, and the combined questionnaire responses plus follow-ups were \textit{comparable in volume to transcribed Zoom interviews}. \textit{We used the \textbf{GPT-4o} model to translate non-English responses into English for analysis, and to translate our follow-up questions into participants' native languages to facilitate communication.} Details of our multilingual communication and translation procedures are provided in Section 3.1.5.

In total, \textbf{8 participants} adopted the questionnaire format (\C{13},\C{14},\C{15},\C{18},\C{20},\C{21},\G{2},\G{5}). \textbf{Zoom} interviews lasted \textbf{35--60 minutes}; audio and text were transcribed with consent, and each participant received \textbf{50 CNY} (about \$7 USD) compensation.

All participants were able to read English, although some preferred to speak or write in other languages. We therefore treated English as the “pivot” language for storage and analysis, while allowing interview and questionnaire interactions in Mandarin Chinese, Japanese, and other languages when this made participants more comfortable.

\subsubsection{Rating Estimation.}
Because Codeforces ratings fluctuate weekly, we report each contestant’s rating as of their interview date. For participants without a stable public rating, we derived an \textit{approximate} rating using: (i) expert assessment by problem setter \Pp{1} (an ICPC/IOI world champion) on anonymized results, and (ii) participant self-assessment against Codeforces tiers. Estimated cases are labeled in Table~\ref{tab:contestants}.

\subsubsection{Interview Guide.}
We designed a semi-structured interview guide aligned with our three research questions (RQ1--RQ3) and tailored to the four stakeholder roles (contestants, problem setters, coaches, and platform stewards). The guide ensured consistent coverage of workflows, fairness, and governance, while leaving flexibility for follow-up questions.

\textbf{RQ1 (Workflows).}
We asked how day-to-day practices had changed since the advent of LLMs, focusing on training, live contests, and post-contest review. Contestants described whether and how they used or avoided LLMs in these stages; setters discussed using LLMs for drafting statements, generating data, or solver checks; coaches reflected on students’ reliance on LLMs and adjustments to training routines; stewards described changes in monitoring and reviewing suspicious submissions. Follow-ups probed concrete episodes and perceived benefits or risks.

\textbf{RQ2 (Fairness).}
We then asked participants to locate the boundary between legitimate assistance and cheating. Contestants outlined what they considered fair or unfair AI use; setters and coaches discussed how they advise others (e.g., whether easy problems should resist LLMs, how to talk to students about acceptable use); stewards described how they decide whether AI-generated solutions violate integrity norms. Follow-ups invited discussion of controversial cases and role-specific dilemmas.

\textbf{RQ3 (Governance).}
Finally, we asked how contest rules and community practices should evolve. Contestants were asked about support for disclosure or tiered restrictions; setters and coaches reflected on platform policies, solver checks, and institutional rules; stewards discussed which policies or tools would make enforcement more credible and how online and onsite contexts differ. Across roles, we blended open-ended prompts with targeted follow-ups to elicit both personal experiences and normative judgments.

\subsubsection{Multilingual communication and translation procedures}

Our core research team is fluent in English, Mandarin Chinese, and Japanese, and has reading proficiency in several other languages represented in the study. Interviews were conducted in English and Mandarin, while long-form questionnaire responses and follow-up emails were received in English, Chinese, Japanese, and a few other languages. Before data collection, we confirmed that all participants could read English well enough to understand the consent form and clarifications. Overall, about 17\% of long-form questionnaire responses contained non-English text.

When we needed to translate non-English text into English for analysis, or to translate follow-up questions into participants’ preferred languages, we used the \textbf{GPT-4o} model strictly as a translation tool. We constrained it with prompts such as: ``Translate the following [source language] text into English. Do not summarize, rewrite, or expand the content beyond translation.'' Raw translations were then reviewed by bilingual members of the research team and, when quoted in the paper, rewritten into our own academic English style. GPT-4o was never used to generate qualitative content, summarize answers, cluster themes, or suggest interpretations; for key excerpts, coders could always consult the original-language text alongside the translation.

In our recruitment materials and consent form, we explicitly asked participants not to use LLMs to write or rewrite their interview or survey answers, while allowing dictionary-style translation tools (including LLM-based translators) to help them express their own views. For long-form responses that appeared ``machine-translated'' in style, we conducted brief follow-up exchanges to confirm that they described participants’ own contest experiences. Although we cannot fully rule out undisclosed LLMs drafting on the participant side, these checks increase our confidence that the qualitative content we analyze reflects participants’ perspectives.

Technical terminology in computing often varies across languages and even across regions within the same language~\cite{Alaofi2024Improving}. 
In one Mandarin interview, for example, a mainland Chinese researcher referred to the concept using the term commonly translated as \textit{binary heap}, while a Taiwanese contestant used a different regionally preferred Mandarin term for the same data structure. 
Although both terms refer to the identical algorithmic concept, the mismatch initially caused brief confusion. 
We resolved this by anchoring the discussion in the canonical English terminology used in programming competition problem statements and standard algorithm textbooks, and recorded the harmonized term in our codebook.

\aptLtoX{\begin{table*}
\centering
\caption{Contestants (C1--C21): countries, ratings, competition years, and gender. Estimated ratings are noted in parentheses. For other stakeholders' table, see the Appendix.}
\label{tab:contestants}
\small
\renewcommand{\arraystretch}{1.1}
\setlength{\tabcolsep}{5pt}
\begin{tabular}{|c|p{0.20\textwidth}|p{0.38\textwidth}|c|c|}
\hline
\cellcolor{myblue}\textbf{Contestant} & \cellcolor{myblue}\textbf{Country or Region} & \cellcolor{myblue}\textbf{Rating} & \cellcolor{myblue}\textbf{Years} & \cellcolor{myblue}\textbf{Gender} \\ \hline
\hypertarget{C1}{\Cid{1}} & China & 3821 (ICPC/IOI World Champion) & 7 & Male \\ \hline

\cellcolor{Lblue}\hypertarget{C2}{\Cid{2}} & \cellcolor{Lblue}Japan & \cellcolor{Lblue}3333 & \cellcolor{Lblue}4 & \cellcolor{Lblue}Male \\ \hline
\hypertarget{C3}{\Cid{3}} & China & 3065 & 7 & Male \\ \hline

\cellcolor{Lblue}\hypertarget{C4}{\Cid{4}} & \cellcolor{Lblue}UK & \cellcolor{Lblue}1858 & \cellcolor{Lblue}4 & \cellcolor{Lblue}Male \\ \hline
\hypertarget{C5}{\Cid{5}} & USA & 2202 & 6 & Male \\ \hline

\cellcolor{Lblue}\hypertarget{C6}{\Cid{6}} & \cellcolor{Lblue}China & \cellcolor{Lblue}2400+ (estimate; Women's National Champion) & \cellcolor{Lblue}5 & \cellcolor{Lblue}Female \\ \hline
\hypertarget{C7}{\Cid{7}} & Hong Kong, China & 2118 & 5 & Male \\ \hline

\cellcolor{Lblue}\hypertarget{C8}{\Cid{8}} & \cellcolor{Lblue}Chinese Taipei & \cellcolor{Lblue}2115 & \cellcolor{Lblue}5 & \cellcolor{Lblue}Male \\ \hline
\hypertarget{C9}{\Cid{9}} & China & 1500 & 3 & Male \\ \hline

\cellcolor{Lblue}\hypertarget{C10}{\Cid{10}} & \cellcolor{Lblue}Egypt & \cellcolor{Lblue}1855 & \cellcolor{Lblue}3 & \cellcolor{Lblue}Male \\ \hline
\hypertarget{C11}{\Cid{11}} & China & 1505 & 3 & Male \\ \hline

\cellcolor{Lblue}\hypertarget{C12}{\Cid{12}} & \cellcolor{Lblue}India & \cellcolor{Lblue}1551 & \cellcolor{Lblue}5 & \cellcolor{Lblue}Male \\ \hline
\hypertarget{C13}{\Cid{13}} & Ukraine & 1391 & 2 & Male \\ \hline

\cellcolor{Lblue}\hypertarget{C14}{\Cid{14}} & \cellcolor{Lblue}Iran & \cellcolor{Lblue}2111 & \cellcolor{Lblue}5 & \cellcolor{Lblue}Male \\ \hline
\hypertarget{C15}{\Cid{15}} & Ukraine & 1592 & 2 & Male \\ \hline

\cellcolor{Lblue}\hypertarget{C16}{\Cid{16}} & \cellcolor{Lblue}Belarus & \cellcolor{Lblue}1191 & \cellcolor{Lblue}1 & \cellcolor{Lblue}Male \\ \hline
\hypertarget{C17}{\Cid{17}} & China & 1000 (estimate; Newbie) & 1 & Female \\ \hline

\cellcolor{Lblue}\hypertarget{C18}{\Cid{18}} & \cellcolor{Lblue}Russia & \cellcolor{Lblue}1250 & \cellcolor{Lblue}2 & \cellcolor{Lblue}Female \\ \hline
\hypertarget{C19}{\Cid{19}} & Brazil & 1617 & 2 & Male \\ \hline

\cellcolor{Lblue}\hypertarget{C20}{\Cid{20}} & \cellcolor{Lblue}Brazil & \cellcolor{Lblue}1229 & \cellcolor{Lblue}1 & \cellcolor{Lblue}Female \\ \hline
\hypertarget{C21}{\Cid{21}} & Brazil & 1009 & 1 & Male \\ \hline
\end{tabular}
\Description{Table of 21 contestants (C1--C21) listing country, Codeforces rating, years of contest experience, and gender. Ratings range from 1000 (newbie) to 3821 (world champion). Most contestants are male; four are female (C6, C17, C18, C20). Participants come from diverse countries including China, Japan, UK, USA, Egypt, India, Ukraine, Iran, Belarus, Russia, and Brazil.}
\end{table*}}{\begin{table*}[t]
\centering
\caption{Contestants (C1--C21): countries, ratings, competition years, and gender. Estimated ratings are noted in parentheses. For other stakeholders' table, see the Appendix.}
\label{tab:contestants}

\small
\renewcommand{\arraystretch}{1.1}
\setlength{\tabcolsep}{5pt}
\rowcolors{2}{ContestantBg!50}{white}

\begin{tabular}{|c|p{0.20\textwidth}|p{0.38\textwidth}|c|c|}
\hline
\rowcolor{ContestantBg}
\textbf{Contestant} & \textbf{Country or Region} & \textbf{Rating} & \textbf{Years} & \textbf{Gender} \\ \hline

\hypertarget{C1}{\Cid{1}} & China & 3821 (ICPC/IOI World Champion) & 7 & Male \\ \hline
\hypertarget{C2}{\Cid{2}} & Japan & 3333 & 4 & Male \\ \hline
\hypertarget{C3}{\Cid{3}} & China & 3065 & 7 & Male \\ \hline
\hypertarget{C4}{\Cid{4}} & UK & 1858 & 4 & Male \\ \hline
\hypertarget{C5}{\Cid{5}} & USA & 2202 & 6 & Male \\ \hline
\hypertarget{C6}{\Cid{6}} & China & 2400+ (estimate; Women's National Champion) & 5 & Female \\ \hline
\hypertarget{C7}{\Cid{7}} & Hong Kong, China & 2118 & 5 & Male \\ \hline
\hypertarget{C8}{\Cid{8}} & Chinese Taipei & 2115 & 5 & Male \\ \hline
\hypertarget{C9}{\Cid{9}} & China & 1500 & 3 & Male \\ \hline
\hypertarget{C10}{\Cid{10}} & Egypt & 1855 & 3 & Male \\ \hline
\hypertarget{C11}{\Cid{11}} & China & 1505 & 3 & Male \\ \hline
\hypertarget{C12}{\Cid{12}} & India & 1551 & 5 & Male \\ \hline
\hypertarget{C13}{\Cid{13}} & Ukraine & 1391 & 2 & Male \\ \hline
\hypertarget{C14}{\Cid{14}} & Iran & 2111 & 5 & Male \\ \hline
\hypertarget{C15}{\Cid{15}} & Ukraine & 1592 & 2 & Male \\ \hline
\hypertarget{C16}{\Cid{16}} & Belarus & 1191 & 1 & Male \\ \hline
\hypertarget{C17}{\Cid{17}} & China & 1000 (estimate; Newbie) & 1 & Female \\ \hline
\hypertarget{C18}{\Cid{18}} & Russia & 1250 & 2 & Female \\ \hline
\hypertarget{C19}{\Cid{19}} & Brazil & 1617 & 2 & Male \\ \hline
\hypertarget{C20}{\Cid{20}} & Brazil & 1229 & 1 & Female \\ \hline
\hypertarget{C21}{\Cid{21}} & Brazil & 1009 & 1 & Male \\ \hline
\end{tabular}
\Description{Table of 21 contestants (C1--C21) listing country, Codeforces rating, years of contest experience, and gender. Ratings range from 1000 (newbie) to 3821 (world champion). Most contestants are male; four are female (C6, C17, C18, C20). Participants come from diverse countries including China, Japan, UK, USA, Egypt, India, Ukraine, Iran, Belarus, Russia, and Brazil.}
\end{table*}}

\subsection{Survey of Global Competitive Programming Participants}
\label{sec:survey}

\subsubsection{Participants.}
To extend beyond interviewees and capture a broader and more representative range of perspectives, we conducted a \textbf{global online survey} that ultimately yielded \textbf{207 valid responses}. Recruitment targeted multiple competitive programming communities, including \textbf{Codeforces}, \textbf{Luogu}, \textbf{Reddit} communities, and \textbf{X}, ensuring outreach to both established and emerging contest regions.

\begin{figure}
    \centering
    \includegraphics[width=1\linewidth]{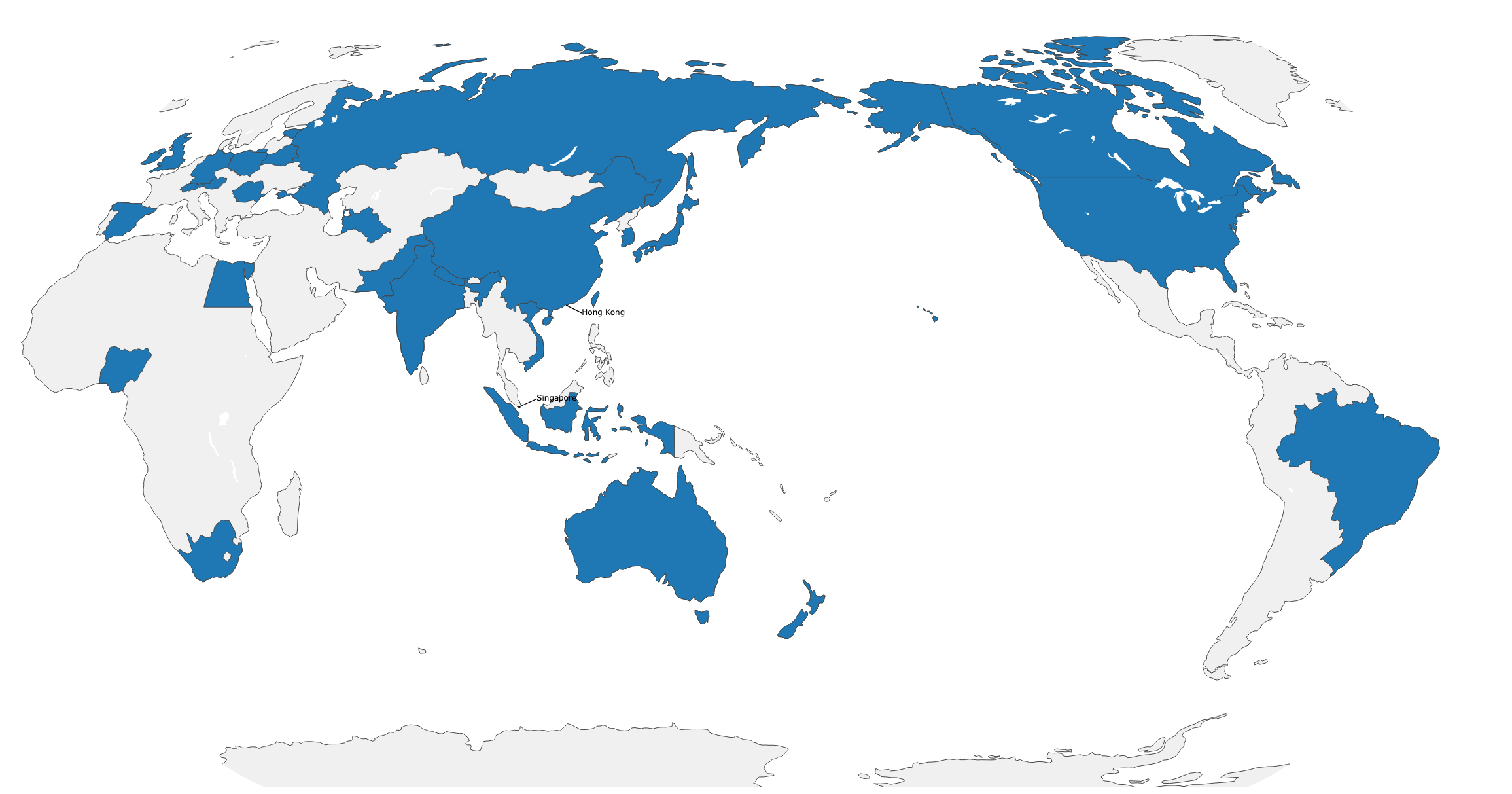}
    \caption{Geographical coverage of survey respondents. Countries and regions with respondents are shown in blue.}
    \Description{A world map showing survey coverage. Countries and regions with respondents are shaded in blue, spanning North America, Europe, Asia, South America, Africa, and Oceania. Small regions that are difficult to see on the map, including Singapore and Hong Kong, are labeled separately to highlight their inclusion.}
    \label{fig:placeholder}
\end{figure}

Prior to analysis, we conducted rigorous \textbf{data cleaning}. We excluded incomplete submissions, responses that were clearly irrelevant to competitive programming, and entries with implausible values. Specifically, exclusion criteria included cases such as reporting impossible contest experience (e.g., ``500 years'' of participation), providing open-ended responses unrelated to programming contests, or indicating non-existent countries or regions.
 For respondents who reported only their Codeforces \textit{tier} (e.g., \textit{Expert}, \textit{Pupil}) instead of a numeric rating, we \textbf{imputed} the median value of the official rating interval (Table~\ref{tab:cf-rating}) to maintain comparability across the dataset. For subsequent analysis, we grouped participants by their self-reported Codeforces rating into three tiers: \textbf{Novices} (Rating$\leq 1399$, $N=59$), \textbf{Intermediates}  (Rating: $1400$–$2099$, $N=116$), and \textbf{Masters+} (Rating $\geq 2100$, $N=32$).

Consistent with the well-documented gender imbalance in competitive programming~\cite{Yamaguchi2024The,Steegh2019Gender}, the final sample was \textbf{predominantly male} (\textbf{190}, 91.8\%), alongside \textbf{15 female respondents} (7.2\%) 
and \textbf{2 non-binary respondents} (1.0\%). Geographically, respondents were highly diverse, spanning \textbf{North America}, \textbf{East and South Asia}, \textbf{Europe}, \textbf{Oceania}, \textbf{Africa}, and \textbf{Latin America}.

\subsubsection{Survey Instrument.}
The survey was structured into four parts: demographics, LLM usage and workflows, fairness and governance, and open-ended prompts. 
Demographics captured background factors such as Codeforces rating, years of experience, gender, and region, which provided context for interpreting responses. 
Workflow items (Likert 1--5) assessed the frequency and perceived impact of LLM usage in practice and preparation, directly addressing \textbf{RQ1}. 
Fairness and governance items (Likert 1--5) probed community norms around cheating, acceptable assistance, and trust in platform policies, aligning with \textbf{RQ2} and \textbf{RQ3}. 
Two open-ended questions asked participants to describe additional LLM-related behaviors and suggest governance measures, allowing us to capture perspectives not covered by predefined items. A minority of open-ended responses were written in languages other than English; these were translated into English using the translation-only GPT-4o procedure described in Section~3.1.5 before coding.

We chose Likert scales to capture gradations of agreement and designed open-ended items to triangulate with interview data. 
A complete overview of all items is provided in \textbf{Appendix Tables~B1--B4}. Reliability checks of the survey showed acceptable consistency (\textbf{Cronbach’s $\alpha=0.76$}), good sampling adequacy (\textbf{KMO=0.80}), and significant Bartlett’s test (\textbf{$\chi^{2}=1788.27$, $df=253$, $p<.001$}), confirming inter-item correlations were appropriate.

\subsection{Data Analysis}

\noindent Our analysis foregrounded the four stakeholder roles in interviews (contestants, problem setters, coaches, platform stewards), while situating them against broader patterns from the global survey of 207 contestants. Interviews served as the primary dataset; the survey provided complementary breadth across \textbf{RQ1 (workflows)}, \textbf{RQ2 (fairness)}, and \textbf{RQ3 (governance)}.

\paragraph{Interview coding.}
We employed \textbf{reflexive thematic analysis}~\cite{Braun2019Reflecting}, iterating through familiarization, open coding, axial coding, and theme refinement. Excerpts were tagged by role, enabling both within-role and cross-role contrasts. The codebook followed a three-level structure: \emph{Level~1 domains} aligned with the RQs; \emph{Level~2 categories} grouped recurring clusters (e.g., training routines, fairness boundaries, enforcement practices); and \emph{Level~3 codes} captured fine-grained practices (e.g., LLMs for templates, AI-solver checks, disclosure norms). Two researchers independently coded an initial subset and refined the scheme via negotiated coding, memoing, peer debriefs, and negative-case analysis. In line with reflexive TA, no inter-rater coefficients were computed. For \textbf{RQ1}, multi-role participants were coded by the role they were speaking from; for \textbf{RQ2} and \textbf{RQ3}, which were not inherently role-specific, excerpts were attributed to each participant’s primary identifier.

\paragraph{Integration.}
In the Findings, interview narratives provide the backbone of evidence across roles, while survey results broaden contestant perspectives. For \textbf{RQ1} and \textbf{RQ2}, survey statistics and open-text responses illustrate usage patterns and fairness norms at scale. For \textbf{RQ3}, survey responses capture contestants’ views on governance, complemented by interviews with setters, coaches, and stewards. This integration balances interview depth with survey breadth while avoiding over-claiming from self-reported, non-experimental data.

\subsection{Platform-Level Statistics on Cheating and Automation}
\label{sec:platform-stats}

To contextualize our fairness and governance findings (RQ2 and RQ3), we analyzed large-scale platform data from Codeforces. Using the official Codeforces API, the second author collected contest metadata, full standings, and complete submission logs for all rated contests held between 2022 and 2025, excluding unrated or special events. The final dataset covers \textbf{336 rated contests}, comprising \textbf{4,844,702 participation instances} and \textbf{64,277,033 submission records} across Div.~1--4, Educational rounds, and Global rounds. For time-based comparisons, we split contests into pre- and post-regulation periods, using the date of Codeforces’ AI policy announcement (\textbf{14 September 2024}) as the breakpoint~\cite{Mirzayanov2024CFPolicy}.

\subsubsection{Official sanction metrics.}
Codeforces marks removed contestants as \texttt{OUT\_OF\_COMPETITION} and invalidated submissions as \texttt{SKIPPED}. Using these fields, we define two indicators, with formal definitions provided in Appendix~\ref{sec:appendix-metrics} (Equations~\ref{eq:severe-rate}–\ref{eq:mild-rate}). The \emph{severe-cheater rate} denotes participants removed due to any \texttt{SKIPPED} submission, while the \emph{mild-cheater rate} denotes participants who remain in the standings but have at least one invalidated submission. These metrics capture officially recorded violations but not harder-to-detect behavioral anomalies.

\subsubsection{Behavioral proxies for automation and AI assistance.}
To detect shifts beyond formal sanctions, we construct two behavioral proxies from Codeforces logs, defined in Appendix~\ref{sec:appendix-metrics} (Equations~\ref{eq:py-share}–\ref{eq:cv}):

\paragraph{(1) Python share.}
The proportion of accepted submissions written in Python or PyPy, which may rise if contestants adopt tool- or model-mediated workflows.

\paragraph{(2) Temporal uniformity (CV)}
For contestants with at least three accepted submissions, we compute the coefficient of variation (CV) of solve intervals. Extremely low CV (CV~$<\!0.2$ or $<\!0.05$) indicates almost clock-like pacing. We aggregate such cases per contest and division as proxies for scripted or automated workflows.

\subsubsection{Steward follow-up.}
Following the formal interview, the first author held multiple informal follow-up conversations with stewards \G{1} and \G{3} to clarify detection workflows, appeals, and perceived false-positive rates. In these exchanges, \G{1} and \G{3} also shared anonymized summaries for ten recent contests on their platforms, including the number of flagged accounts, sanctions, and successful reversals. These practitioner insights and logs helped us interpret the platform-level statistics, particularly the discrepancy between stable official sanction rates and the rising prevalence of behavioral anomalies in the logs.

We use these metrics descriptively in the Findings to triangulate interview and survey evidence on fairness, AI-assisted cheating, and the evolving limits of automated detection.

\section{Results}

\textit{In this section, we used GPT-4o to assist with writing Python scripts. It was employed to clean the raw data and to generate Jupyter Notebook code for data visualization. GPT-4o did not intervene in the selection of data analysis methods.}

\subsection{RQ1 – Workflow: How are LLMs integrated into stakeholders’ workflow?}\label{sec:contestants-workflow}

\subsubsection{Contestants}
This subsection examines how contestants integrate LLMs into their workflows across four phases of contest activity—pre-contest setup, in-contest performance, post-contest review, and daily training—drawing on 37 interviews and a survey of 207 participants.

\textbf{Pre-contest setup.} Only a few contestants reported using LLMs right before a contest, and those uses were pragmatic and infrequent. \C{5} asked AI tools to generate or adapt boilerplate for \emph{segment tree} or \emph{union–find} in their own coding style as a one-off setup; \C{9} used LLMs to translate archived problems or to refresh advanced concepts. As a top-rated competitor \C{2} added, \emph{``I also ask it about PC environment setup and how to use programming languages.''} In practice, this “pre-contest” use meant tidying the editor/compilation setup, translating a few materials, and lining up trusted templates—not rehearsing solutions. Consistent with this light-touch pattern, survey respondents rated “LLMs are a regular part of my workflow” relatively low (\emph{M}=2.31, \emph{SD}=1.34, 1–5), and few reported having \emph{changed} preparation because of LLMs (\emph{M}=2.32, \emph{SD}=1.32, 1–5).

\textbf{In-contest behavior.} Contestants largely avoid AI tools, citing strict platform bans, contest spirit, and doubts about usefulness under time pressure. As one top competitor noted, \emph{“using ChatGPT mid-contest never even crosses my mind.”} The only exceptions were occasional statement translation or light IDE autocomplete. Survey results echo this norm, with strong agreement that AI use is cheating (\emph{M}=4.56) and unfair (\emph{M}=4.45), paired with high willingness to comply with rules (\emph{M}=4.55). Still, several acknowledged awareness of peers who cheated, a concern also raised in open-ended responses.

\textbf{Post-contest review.} After contests, LLMs were routinely used for upsolving—asking for hints, simplified explanations, or editorial-style guidance. Contestants noted faster learning (\C{11}) and time savings (\C{10}), with others reporting deeper conceptual gains (e.g., Bellman–Ford, sparse tables, counterexamples, runtime errors). Survey results echo this pattern: LLMs were seen to aid learning new techniques (\emph{M}=3.15) and provide modest gains in efficiency (\emph{M}=2.77).

\textbf{Daily training.} Outside contests, LLMs were used as “rubber-duck” tutors—providing hints, debugging help, or rephrasings to clarify concepts (\C{6}, \C{11}). Contestants also refreshed knowledge, checked syntax (\C{9}), or drafted templates for common data structures (\C{5}, \C{9}, \C{12}), reducing boilerplate and focusing on problem-solving. Survey results show moderate usage (\emph{M}=2.48), with 43\% using LLMs at least occasionally and 25\% often; perceived performance gains were limited (\emph{M}=2.15).

\textbf{Risks and boundaries.} Training with AI was not without risks. Contestants highlighted reliability issues—generated code \emph{“often fails on edge cases”} \C{14}; models sometimes produce confident but wrong explanations \C{4}—and worried about dependency: \emph{“if I let it do all the thinking, I won’t improve my own skills,”} \C{7}. These concerns were broadly shared in the survey (\emph{M}=3.71, \emph{SD}=1.44, 1–5). Normative boundaries remained firm \emph{in-contest} (cheating/unfairness means above 4.45), even among those who used LLMs frequently during practice.

\begin{figure}[h]
  \centering
  \includegraphics[width=0.85\linewidth]{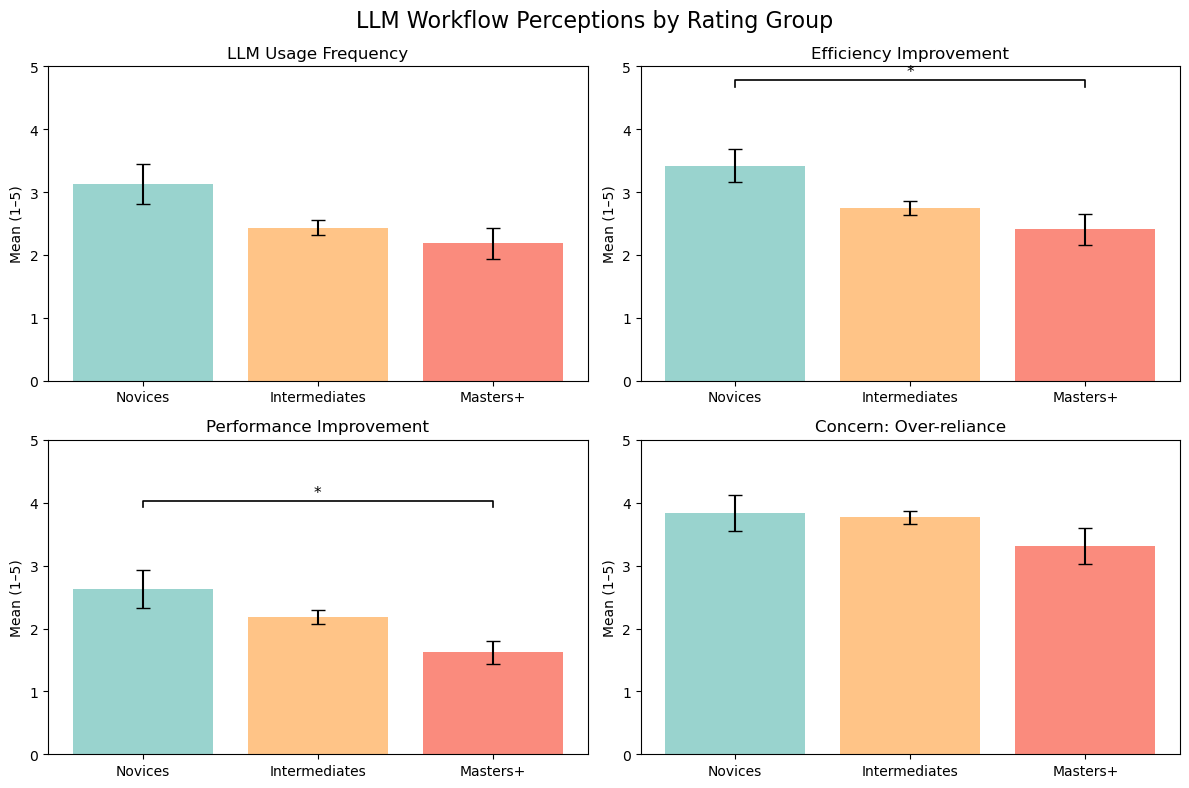}
  \caption{Perceptions of LLM usage across rating groups. Bars show means with SEM. 
  Pairwise Mann–Whitney U tests (Bonferroni corrected) indicate that novices perceive 
  significantly greater efficiency and performance gains than Masters+ ($p < .05$), 
  whereas concerns about over-reliance show no significant differences.}
  \Description{Bar charts comparing three groups of programming contest participants 
  (Novices, Intermediates, Masters+) across four measures: frequency of LLMs use, perceived 
  efficiency improvement, perceived performance improvement, and concern about over-reliance. 
  Novices report higher frequency, efficiency, and performance gains, while all groups show 
  similar concern about over-reliance. Error bars indicate standard error of the mean.}
  \label{fig:contestant_skill_gradient}
\end{figure}

\textbf{Adoption gradient by skill.} Both interviews and survey data show a clear gradient: novices experiment most, intermediates adopt selectively, and elites avoid LLMs. Usage was inversely correlated with rating ($\rho=-0.227$, $p=.001$), with novices reporting greater efficiency and performance gains, while Masters+ remained the most restrained. Concerns about over-reliance were shared across groups without significant differences. As \C{2} noted, relying on models to fill reasoning gaps was “mostly a waste of time,” reflecting elite skepticism toward core problem-solving use.

In general, contestants’ LLM use followed a four-phase rhythm: little in pre-contest setup, almost none in contests, and much heavier reliance in post-contest review and daily training. They valued LLMs for clarifying concepts, debugging, and offloading boilerplate, but noted unreliability on edge cases and limited help with the hardest problems. Novices used LLMs as accelerators, while elites kept them at arm’s length. Overall, self-reports depict LLMs as embedded in learning but excluded from live contests—even as many acknowledged peers who cheated.

\subsubsection{Problem Setters}

\begin{figure*}[t]
  \centering
  \includegraphics[width=\textwidth]{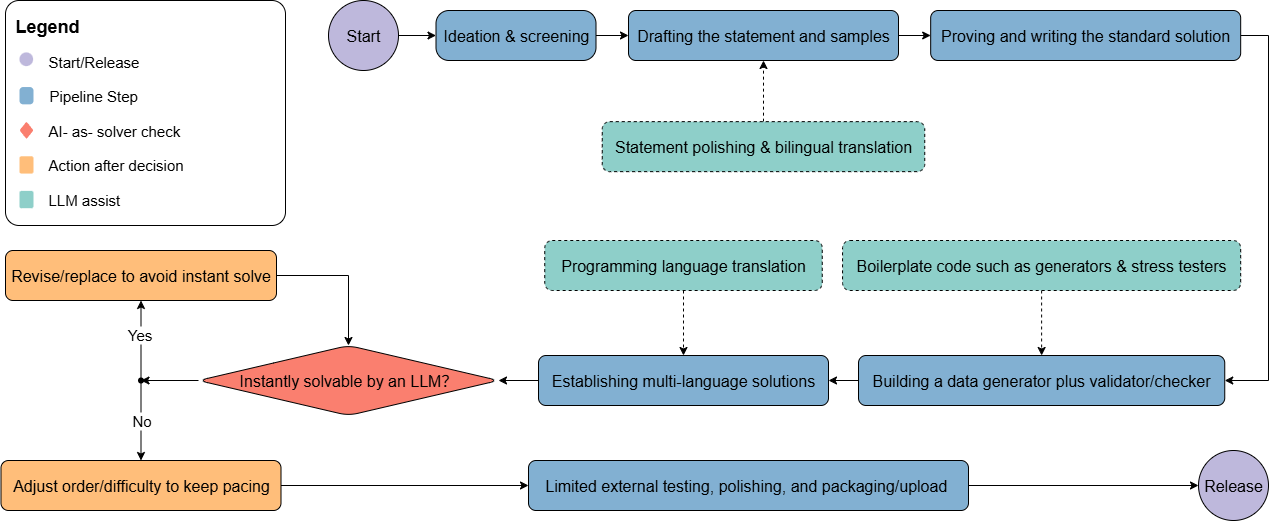}
  \caption{Workflow change of problem setters. 
  While the ideation and screening stage remains fully human-driven, subsequent steps of the pipeline (drafting, proofing, solution writing, validation, translation, and polishing) increasingly integrate LLM assistance for low-creativity and repetitive tasks. Importantly, setters emphasized that AI should never intrude into the creative ideation phase, as it undermines the originality and aesthetic value of competitive programming problem design.}
  \Description{A flowchart of problem setter workflow. It starts with human-only ideation and screening, followed by drafting statements, writing standard solutions, and building data generators. LLM assist is introduced in polishing statements, bilingual translation, language translation, and boilerplate code such as stress testers. An AI-as-solver checkpoint checks if the problem is instantly solvable by an LLM. If yes, problems are revised or replaced; if no, setters adjust order and pacing. The process ends with external testing, polishing, packaging, and final release. The chart highlights that AI is excluded from ideation but accepted in routine tasks.}
  \label{fig:problemsetter_workflow}
\end{figure*}

Competitive programming problem setting is notably “self‑recruiting”: almost all setters are current or former contestants; in our sample, all seven interviewees fit this pattern, and \Pp{2-4}—are still actively competing while setting.

\emph{Pre‑LLMs baseline workflow.} Across interviews we observed a stable, six‑stage pipeline: (i) ideation and screening (avoid near‑duplicates, assess novelty and pedagogical value); (ii) drafting the statement and samples; (iii) proving and writing the standard solution  (iv) building a data generator plus validator/checker; (v) establishing multi‑language solutions (e.g., C++→Python/Java); and (vi) limited external testing, polishing, and packaging/upload before release (\Pp{1-7}).

\emph{Where LLMs now sit in the pipeline.} Setters do not delegate ideation to LLMs, but many fold models into low‑creativity, repetitive work: (a) \textbf{AI‑as‑solver checks} before release to gauge whether a task is “instantly” solvable by a model—most commonly for easy/check‑in tasks (\Pp{1},\Pp{7}); some extend this to hard problems to ensure that the problem order and contest dynamics remain stable (\Pp{3}). (b) \textbf{Boilerplate code} such as random test generators and stress testers is frequently drafted by LLMs and then reviewed by humans (\Pp{2},\Pp{3}). (c) \textbf{Programming language translation} (e.g., C++→Python/Java) is used to produce required standard solutions(\Pp{1},\Pp{3}). (d) \textbf{Statement polishing and bilingual translation} helps to explain the problem statement more clearly (\Pp{1,3,6}). By contrast, \Pp{5} keeps core implementations fully hand‑written.

\emph{Divergent stances on “AI‑proofing” easy tasks.} Two positions emerged. One camp argues that preventing AI from solving easy tasks is \emph{not} the setter but platform steward’s job(\Pp{3},\Pp{5}). The other camp treats AI resistance as a release criterion for easy or teaching contests: if a model can instantly solve a task, it should be replaced with another one(\Pp{2,4,7}).

\emph{Pacing risk on hard problems.} A recurring concern is scoreboard stability. One setter emphasized that even hard problems should undergo an AI‑as‑solver pass: if a model can crack a top task early, humans may chase it and neglect solvable items, degrading contest quality (\Pp{3}). Relatedly, the former \textbf{world champion}(\Pp{1}), served as an offline ICPC finals setter/judge reported an air‑gapped event where a world‑leading company fielded a non‑public model as a “team.” To guard early pacing, organizers negotiated a two‑hour constraint to attempt only problems already solved by humans. The model ultimately only solved 1 out of 13 and finished near the bottom, yet the setter remains cautious about future model capabilities. This episode reinforces the value of pre‑release AI checks for contest pacing. 

\emph{Convergences.} Despite stylistic diversity, several patterns are consistent. First, \textbf{no AI for ideation}: participants view model‑generated “ideas” as undermining originality and the craft of problem design (\Pp{1-7}). Second, \textbf{LLMs are welcome in low‑creativity slots}: brute/stress tooling, generators/validators, cross‑language references, and statement polishing (\Pp{1-4,6}). Third, \textbf{AI‑as‑solver testing is normalizing} as a lightweight checkpoint, but it is not always a hard gate; gate strength depends on the contest round’s purpose (\Pp{2,4,7}). Finally, \textbf{easy problems are shifting from “template application” to “short reasoning”}: setters increasingly encode small abstractions or gentle twists to preserve novice value while dampening model shortcuts (\Pp{1,6,7}).

Overall, LLMs have entered nearly every phase of the problem-setting pipeline, offering automation, translation, and verification across low-creativity tasks. While they do not replace human creativity, they reshape how setters allocate their time and attention—introducing AI-as-solver checkpoints, expanding generator and validator automation, and supporting bilingual clarity. These shifts are not merely peripheral: they recalibrate workflow priorities and blur boundaries between technical assistance and judgment.

\subsubsection{Coaches} 
Across interviews and prior studies, the influence of coaches in competitive programming training varied greatly by region~\cite{liu2008training}. In many countries, coaches mainly fulfilled administrative functions~\cite{park2007roles}, while in China, Russia, and the United States they were more directly involved in novice instruction and early team selection. The \textit{baseline workflow} before LLMs typically involved coaches giving introductory lectures and weekly practice sessions for interested students, organizing school-level contests on weekends, and—when representing the university in national or international competitions—selecting teams according to rules they established \T{1-7}. As one highly experienced coach reflected, \T{3}: \emph{“Even though I have over twenty years as a contestant and more than a decade as a coach—and have led teams to World Finals medals—I no longer feel capable of teaching contestants above Codeforces 2000. My role remains in nurturing novices and selecting promising candidates.”} All interviewed coaches also shared a common understanding that advanced contestants primarily improve through self-study with community resources, peer learning, and active participation in online and onsite contests; targeted coaching for advanced players is extremely scarce.

With the rise of LLMs, coaches noted that their novice-oriented instruction has remained largely stable, with only minor adjustments—such as occasionally asking an LLM to review long-used lecture slides to check whether the material feels outdated in a rapidly evolving field (\T{3}). Beyond this light use, they emphasized that the pedagogical core of their teaching had not been altered (\T{1,5,6,7}). The most visible changes instead arose in the \textit{selection workflow}, where universities increasingly confronted candidates suspected of relying on LLMs: first offering pedagogical warnings, but excluding persistent offenders (\T{2,3,4}). As one coach stressed, \T{4}: \emph{“We cannot allow students with academic dishonesty to represent our university.”} At the same time, coaches such as \T{7} reminded that “the competitive programming community itself is built on norms of fairness and justice; any form of illegitimate advantage immediately sparks controversy.” Taken together, coaches agreed that apart from integrity enforcement and light material checks, their core instructional practices remained largely unchanged.

\subsubsection{Platform Stewards} 
Across interviews, stewards agreed their workflows now center on monitoring and detection, though strategies diverged: some (\G{1,2}) described proprietary systems, usually limited to entry-level contests, while others (\G{4}) relied on plagiarism checks due to cost. All noted that triaging AI-related reports has become a major new burden, and community-driven oversight initiatives (\G{7,8}) now complement official enforcement. This hybrid landscape of formal and grassroots governance raises broader debates about enforcement, cost, and trust, which we return to in Section~\ref{sec:4.3}.

\subsubsection*{Answer to RQ1}
LLMs have reshaped day-to-day workflows across all four stakeholder groups in distinct ways. 
Contestants reported minimal use before contests, near-complete avoidance during contests, and much heavier reliance afterward and in daily training. They used LLMs to clarify concepts, debug, and offload boilerplate, while acknowledging limits on edge cases and hardest problems. Novices leaned on LLMs as accelerators and tutors, whereas elites deliberately minimized use to preserve independent problem-solving.  

Problem setters kept the established six-stage pipeline but integrated LLMs into low-creativity work such as solver checks, generator scripts, template code, and bilingual polishing, while reserving ideation and contest pacing for humans.  

For coaches, core instructional practices remained stable. The main change was in selection workflows, where they increasingly screened for AI misuse and excluded persistent offenders, while occasionally using LLMs to update long-standing materials.  

Platform stewards shifted most sharply: from system stability toward monitoring, detection, and triaging reports. Some adopted proprietary detection systems, others relied on plagiarism checks and contestant self-discipline, but all described report handling as a major new burden. The rise of independent oversight initiatives further distributed governance across official and grassroots actors.  

Overall, LLMs are now embedded in preparation, training, and infrastructure routines, while being self-reported as largely excluded from live contest performance.

\subsection{RQ2 – Fairness: Perceptions of ``Cheating'' and Fair Use Boundaries}

\subsubsection{Official Rules: Platform Policies and Enforcement}

Interviews with six stewards (\G{1}–\G{6}) and current rule texts reveal a convergent stance: during contests, pasting statements into external models or submitting model-generated code is treated as plagiarism. Allowances are narrowly limited to non-derivative utilities such as statement translation or very light IDE autocompletion, while hints, debugging, or iterative patching are explicitly excluded. Rules are surfaced via compulsory notices at registration, with enforcement layered across automated heuristics (sometimes aided by “bait prompts”), similarity checks, manual review of borderline cases, and community reports. Sanctions are steep and graduated—initial warnings and short suspensions escalating to long or permanent bans, with appeals in AI cases rarely upheld. Against this mainstream, \G{2} described a tiered approach: lower divisions remain AI-free, while top divisions permit tightly bounded use for low-creativity tasks but still forbid outsourcing algorithmic reasoning or full solutions. Across both approaches, the shared principle is clear: problem solving must remain the contestant’s own work.

\subsubsection{Gray Zones: Stakeholders’ Perceptions of Fair Use}

Despite formal anti-cheating policies, all stewards we interviewed acknowledged that online contest rounds cannot be made perfectly clean. Technical, operational, and ethical constraints (from sophisticated AI outputs that evade detection to the impracticality of intrusive proctoring) make airtight enforcement impossible (\G{1}--\G{6}). As \G{6} remarked, \emph{``since LLMs arrived, absolute fairness is gone, so we can only aim for relative fairness.''} In short, today’s online contests contain substantial \textbf{gray zones} that rules alone cannot eliminate, and rule-makers are continually updating policies as technology and practices evolve.

\begin{figure}[h]
  \centering
  \includegraphics[width=1 \linewidth]{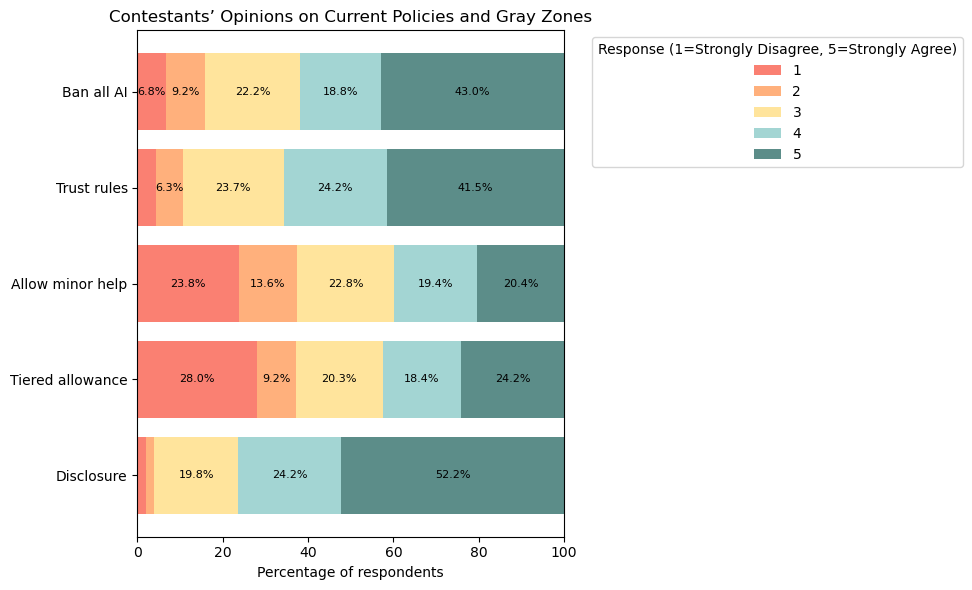}
  \caption{Contestants’ opinions on current policies and gray zones. 
  Stacked bars show Likert distributions (1=Strongly disagree, 5=Strongly agree) 
  for five statements: banning all AI use, trusting current rules, allowing only 
  minor help, adopting tiered allowances, and requiring disclosure. Results show 
  broad support for banning AI and mandatory disclosure, general trust in existing rules, 
  but sharp polarization on minor help and tiered allowance.}
  \Description{Stacked horizontal bar chart comparing survey responses to five 
  policy-related items. Each bar is divided into five segments corresponding to 
  strongly disagree, disagree, neutral, agree, and strongly agree. Most respondents 
  support banning all AI use and requiring disclosure, and many trust existing rules. 
  Responses to allowing minor help and to tiered allowance are split nearly evenly, 
  illustrating contested gray zones. Percentages are labeled within each bar segment.}
  \label{fig:gray_zones}
\end{figure}

Survey results in Fig.~\ref{fig:gray_zones} illustrate these tensions. 
On the ``ban'' axis, \textit{Ban all AI} drew strong support (43\% strongly agree), 
and many respondents expressed confidence in current rules (42\% strongly agree); 
yet over one-third remained unconvinced about their fairness. 
Several open-ended responses bluntly noted that, 
\emph{``since it is technically impossible to detect who illegally uses LLMs, 
the contest rules are essentially self-deception.''} 
Opinions fractured when it came to gray zones: 
\textit{Allow minor help} (e.g., translation/autocomplete) and \textit{Tiered allowance} 
(lower divisions AI-free, upper divisions partially allowed---already practiced by \G{2}) 
each showed nearly even splits between support ($\approx40$\%) and opposition ($\approx37$\%). 
By contrast, a clear majority favored mandatory \textit{Disclosure} if AI was used (76\% agree/strongly agree).

To probe how such gray zones might be understood in practice, we developed five short vignettes from three sources: (i) gaps in policy wording, (ii) practices repeatedly described in interviews, and (iii) self-reports from survey items. Using reflexive thematic analysis, two researchers refined these into five scenarios and presented them to 37 interviewees for judgment. Each was asked whether it should be allowed, forbidden, or judged “it depends.” The scenarios were: (1) translating statements, (2) AI-based autocompletion, (3) generating boilerplate code, (4) pre-submission checking with LLMs, and (5) consulting LLMs for conceptual hints. Stewards acknowledged that none of these can be reliably detected in live contests, raising the question of whether—if discovered—they should be punished or tolerated.

Translation and light autocompletion were generally tolerated, with nearly all respondents considering statement translation acceptable and most treating autocomplete as a continuation of IDE support, though some warned that advanced editors may blur into AI-generated logic. By contrast, template code generation was divisive: a few likened it to reusing libraries, but most viewed it as cheating, accepting only pre-contest preparation. Pre-submission code checking drew universal condemnation, as it bypasses the judge’s role and leaves little trace, though stewards admitted it is nearly impossible to prevent in online rounds.

Conceptual consultation with LLMs was mostly judged a violation, though a handful of elite contestants felt it acceptable when struggling with “final boss” problems, reflecting how their impressions of contests center on solving the hardest tasks. One steward (\G{2}) echoed this permissive view—“if it is only discussion, why not?”—yet paradoxically was also the only interviewee to advocate legal action against cheaters. Despite such exceptions, the majority still regarded in-contest conceptual hints as a breach of fairness.

\textbf{Summary.} Stakeholders converged at the extremes but split in the middle. Translation and narrow autocomplete were mostly tolerated; pre-submission checking and conceptual consultation were universally condemned; and template generation divided opinion, with most rejecting in-contest use but some accepting pre-contest preparation. What counted as cheating hinged not only on detectability but also on whether AI was seen to replace a contestant’s own reasoning. These contested gray zones set the stage for Section~4.2.3, where we examine more overt and systematic AI-assisted cheating.

\subsubsection{The Dark Side of Workflow: Hidden Misuse and Cheating Practices}

Building on Section~4.2.2’s map of tolerated utilities and contested gray zones, our survey establishes a stark normative baseline: nearly nine in ten respondents agreed that using AI during an official contest is “essentially cheating” (\emph{M}=4.56, \emph{SD}=0.90; 89\% agree/strongly agree) and that allowing LLMs would confer an unfair advantage (\emph{M}=4.45, \emph{SD}=0.99; 86\%). At the same time, while most profess willingness to comply with organizer rules (\emph{M}=4.55, \emph{SD}=0.83; 88\%), confidence in enforcement is notably weaker (trust in fair enforcement $\approx$61\%; satisfaction with current handling $\approx$51\%). This gap between \emph{norms on paper} and \emph{perceived enforceability} sets the stage: beyond self-reported abstention and policy text, interview accounts point to a darker undercurrent of AI misuse that exploits hard-to-observe practices and low detection risk. We now turn from gray zones of “acceptable help” to concrete tactics through which contestants covertly leverage LLMs to gain an illicit edge.

Beneath the survey’s near-consensus that LLMs have little effect on in-contest workflow, interviews and community accounts reveal a parallel story: the emergence of a covert cheating workflow that blends old practices with new tools. Contestants, coaches, setters, and stewards all described how traditional shortcuts—copying code from friends, sharing via email or USB sticks, or buying solutions in Telegram groups—have been supercharged by generative AI.

One contestant (\C{1}) admitted that the “opening move” for some peers is simply to dump every problem statement into an LLM at the start of the contest, asking the model to parallelize the reading and analysis while they wait. This shifts the bottleneck from comprehension to filtering, as cheaters skim AI outputs for anything directly compilable. Others noted cases of users who “throw the whole task to GPT, delete the comments, and paste the result,” laundering the submission with superficial edits. Coaches echoed this pattern: \T{4} described catching students who could not explain their own code, with giveaway traces of AI such as verbose English variable names and over-detailed comments. When those traces were stripped away, detection became more circumstantial—sudden leaps in performance or uncanny similarities across classmates. As \T{2} observed, students who struggled for an hour suddenly solved three or four hard problems in the final minutes of practice, almost certainly with “extra help running quietly in the background.”

Problem setters and stewards described complementary perspectives. \C{5}, coordinating beginner divisions, openly uses AI to pre-generate brute force baselines and stress testers, but also to generate sample “AI solutions” so that suspicious patterns can be spotted on the scoreboard (for example, accounts solving the hardest problem but skipping easier ones). Yet the automatic plagiarism checker that once caught copy–paste now fails against the stylistic diversity of LLMs: \C{5} had seen users rely on AI across thirty contests without a single automated flag. Manual review, cross-checking variables, code structure, and even overengineered solutions for easy tasks, has become the only line of defense. \G{1}, an official steward, described the frustration of scanning thousands of submissions after each Luogu monthly contest: “we can catch two hundred AI cheaters in a single evening, yet next week they come back with new accounts.” Likewise, \T{7}, a volunteer steward, pointed to the growing role of off-platform coordination. Large Telegram groups, in particular, serve as anonymous marketplaces for leaked or AI-generated solutions. Their end-to-end encryption and pseudonymous identities make them exceptionally difficult to trace, leaving organizers with little visibility into the scale of collusion. Community moderators recalled sting operations in which deliberately fragile solutions were leaked; when an obscure edge case triggered, hundreds of synchronized failures revealed the existence of an AI-rewritten supply chain operating beyond official oversight.

These accounts converge on a coherent cheating pipeline. It begins with parallel prompting—feeding all problems to models at once. During the contest, contestants either copy model output directly or funnel marketplace solutions through LLMs to evade plagiarism checks. Post-processing includes laundering (removing comments, renaming variables) and just-in-time editing. Distribution is coordinated in Telegram groups, where anonymity and AI rewriting have replaced traditional copy–paste sharing. Finally, when challenged, cheaters invoke plausible deniability: claiming their code was “pre-generated” before the contest or that AI was used only for translation. As one steward put it, “they know exactly how far to push the gray zone—enough to hide, never enough to admit.”

In contrast to the survey’s finding that most contestants report “little or no” contest-time LLMs use, the hidden layer reveals a systematic repurposing of generative AI into cheating pipelines. The result is a bifurcated reality: publicly, LLMs appear peripheral to the contest workflow; privately, they underpin an evolving underground economy of automation, laundering, and distribution that sustains large-scale misconduct. This gap between the visible and the hidden marks the true dark side of workflow.

Compared to long‑standing forms of cheating such as copying solutions from friends or purchasing code in Telegram groups, LLM‑assisted cheating was seen as both a continuation and an escalation. Participants noted that traditional collusion already undermined fairness, but required human intermediaries and tended to leave recognizable stylistic traces. In contrast, LLMs make it easier to “launder” purchased or shared code into many stylistically distinct variants, lowering the effort and expertise needed to evade plagiarism detectors. At the same time, several stewards emphasized that, normatively, contestants regarded LLM‑based and non‑LLM‑based cheating as equally unacceptable; what changed was not the moral evaluation, but the scale, subtlety, and deniability of misconduct.

\subsubsection*{Answer to RQ2}
Stakeholders largely agreed that using LLMs in live contests is cheating, but opinions split in the gray zones. 

\emph{Official rules}: Platforms ban model-generated code or hints, with only minor allowances such as translation or light autocomplete. Enforcement combines automated checks, plagiarism detection, manual review, and community reports, yet organizers admitted detection is never perfect.  

\emph{Gray zones}: Translation and narrow autocomplete were tolerated, pre-submission checks and conceptual consultation were condemned, and template generation was accepted only in pre-contest preparation.  

\emph{Hidden misuse}: Interviews revealed covert workflows—batch prompting problems, laundering outputs, and distributing rewritten code via encrypted groups—showing how LLMs fuel underground cheating economies.  

Overall, RQ2 exposes tensions between written rules and enforceability, and between tolerated utilities and hidden misuse. The core principle holds that problem solving must remain the contestant’s own work, but fairness is continually contested in practice.

\subsection{RQ3 – Governance: Awareness, Compliance, and Trust}
\label{sec:4.3}

\subsubsection{Official Measures and Contestant Feedback}

Survey responses on governance reveal a consistent compliance–trust gap. Nearly nine in ten participants expressed willingness to follow whatever rules organizers set (\emph{M}=4.55, \emph{SD}=0.83; 87.9\%), and more than three-quarters supported mandatory disclosure of any AI use after contests (\emph{M}=4.23, \emph{SD}=0.96; 76.3\%), suggesting a strong readiness to accept and even tighten formal requirements. Yet confidence in governance was markedly weaker: only about 61\% trusted organizers to enforce rules fairly (\emph{M}=3.71, \emph{SD}=1.11), and just half were satisfied with current handling (\emph{M}=3.44, \emph{SD}=1.14). Awareness of platform policies remained moderate (63.6\%), and while roughly 70\% agreed that rule-making is transparent and inclusive, only 36.9\% had ever participated in related community discussions, indicating that most contestants experience governance as something done \emph{for} them rather than \emph{with} them. Together, these findings portray a community that is normatively compliant but epistemically cautious: contestants broadly accept organizer authority, yet question whether enforcement is capable, consistent, and sufficiently open to grassroots input.

\begin{figure}[t]
    \centering
    \includegraphics[width=0.65\linewidth]{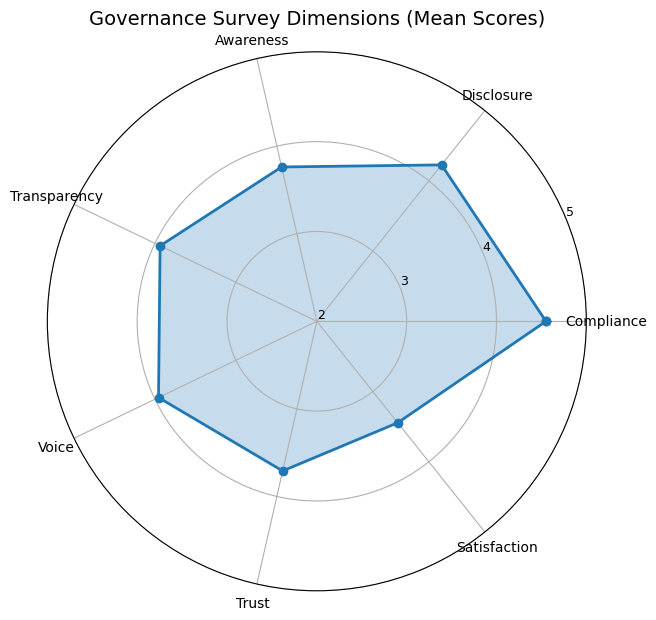}
    \caption{Radar chart of governance-related survey items. Dimensions include \textit{compliance with organizer rules}, \textit{mandatory disclosure}, \textit{policy awareness}, \textit{transparency}, \textit{voice in rule-making}, \textit{trust in enforcement}, and \textit{satisfaction with handling}. High compliance and support for disclosure contrast with weaker trust and satisfaction, revealing a clear compliance–trust gap.}
    \Description{A radar chart showing seven governance dimensions with mean scores on a five-point scale. Compliance and disclosure are highest, transparency, awareness, and voice are mid-range, while trust and satisfaction are lowest, highlighting a compliance–trust gap.}
    \label{fig:radar}
\end{figure}

Platform-level statistics from \textbf{Codeforces} and from the platform stewarded by \G{1} further illuminate this gap. Using the severe-cheater rate $R^{\mathrm{severe}}_c$ from Equation~\ref{eq:severe-rate}—the percentage of contestants removed from the standings because at least one submission is marked \texttt{SKIPPED}—the second author’s analysis shows that severe sanctions are rare and strongly stratified by division: around 2--3\% of participants are removed in Div.~3 and Div.~4, about 1.8\% in Div.~2, and only 0.18\% in Div.~1. Taking 14 September 2024 (when Codeforces announced explicit AI-related policies) as a breakpoint, we did not observe a spike in $R^{\mathrm{severe}}_c$; post-regulation values even decreased slightly in Div.~2 and Div.~3. Internal reports for ten recent contests on \G{1}’s platform show a similar pattern: labelled cheaters typically constitute under 2\% of entrants, appeals are rare, and confirmed misclassifications are vanishingly uncommon, with only a handful of reversals across thousands of sanctioned cases (a misclassification risk well below 0.01\% for normal participants). This confirms stewards’ perception that false positives are almost nonexistent, but also suggests that official logs capture only a narrow slice of misconduct.

By contrast, behavioral proxies derived from Codeforces logs Python share $\rho^{\mathrm{py}}_c$ (Equation~\ref{eq:py-share}) and temporal uniformity $CV(u,c)$ (Equation~\ref{eq:cv})—reveal substantial shifts in how contestants solve problems. After the introduction of AI policies in late 2024, the Python share $\rho^{\mathrm{py}}_c$ (the fraction of accepted submissions written in Python or PyPy) increased markedly across divisions, with large relative gains in mid- and high-level contests despite C++ remaining dominant. Over the same period, the number of contestants whose accepted-submission timestamps exhibited highly regular intervals ($CV(u,c) < 0.2$, i.e., almost clock-like timing) grew sharply in Div.~2, Educational rounds, and Div.~4, with tens of thousands of extreme cases at $CV(u,c) < 0.05$ concentrated in mid-level divisions. These patterns cannot be taken as direct proof of AI-assisted cheating for specific individuals, but they are consistent with the spread of scripted or tool-mediated workflows that often evade formal sanctions.

Against this backdrop, platforms employ a layered ``detect--review--sanction'' pipeline: rules reiterated at registration, similarity checks and AI-style heuristics (including \emph{bait prompts}) during contests, and post-hoc moderator review with penalties up to permanent bans, with appeals rarely succeeding (\G{1}). In entry-level rounds, \G{1} reported hundreds of AI-assisted submissions flagged in a single evening, though easy re-registration weakens deterrence. Stewards noted that proprietary detectors mainly cover beginner divisions, while higher tiers still depend on plagiarism tools and manual review due to cost and fragility. In unproctored settings, perfect detection is impossible; integrity instead relies on self-discipline, public listings of violators, and community oversight. Taken together, contestants are willing to comply with rules and, in practice, very few are mistakenly punished, yet AI-assisted and scripted workflows spread far beyond what official sanctions reveal. Governance therefore operates under a dual constraint: it must maintain participants’ trust in the rarity of false positives while acknowledging that current detection pipelines see only the tip of the iceberg.

\subsubsection{Community Oversight and Grassroots Initiatives}

Beyond official pipelines, \emph{community co-governance} has emerged with an evidence-first, largely manual \emph{report–review–label} workflow. {\Gg{7}} described a browser extension and database where users submit reports with handles and evidence, reviewed by a small moderator team before hourly updates. Automation was minimized after intolerable false positives from early detectors; as {\Gg{7}} noted, “you can’t have a tool that outs people and still be wrong.” Safeguards include ownership checks and second-review confirmation. A cluster of false positives triggered by an LLM idiom (\texttt{unordered\_map.reserve}) was later removed, underscoring the value of reversible labeling.

One striking case involved an interviewee who had earlier denounced AI-assisted cheating, yet was later flagged by a grassroots system with full evidence (linked submissions, pattern analysis, and second-review confirmation). In a brief follow-up conducted after the flag—reported here with the participant’s explicit consent—the contestant explained that the incident occurred during an Educational round solved in a school computer lab. The participant stated that ideas were “borrowed” from nearby classmates while working through the problems, but denied any direct use of LLMs. The participant described the AI-style label as “unlucky but reasonable,” chose not to appeal, and simply opened a new account. In the participant’s view, current detection systems are “unlikely to punish the wrong person, but unlikely to catch everyone,” illustrating both the limitations of self-reported accounts (e.g., social-desirability constraints) and the partial coverage of grassroots review compared to official pipelines.

Grassroots groups also run \emph{adversarial probes}. {\Gg{7}} described “sting” operations seeding fragile solutions into Telegram groups; synchronized failures revealed a model-rewritten supply chain. In parallel, {\Gg{8}} pre-generates exemplar AI solutions to anticipate suspicious patterns (e.g., solving the hardest task while skipping easier ones). Because MOSS-style checks struggle with diverse outputs, human heuristics—such as over-engineered solutions to easy tasks, odd comment cadence, or template fingerprints—remain indispensable.

Accuracy, however, is fragile. {\Gg{7}}’s false-positive case required re-review, and thresholds vary: {\Gg{8}} contrasted single-review inclusion with multi-coach deliberation that cut false positives “very close to zero.” Automation remains limited by detector errors and MOSS baselines, and access is constrained by platform rate caps and encrypted Telegram groups. Even so, interviewees emphasized that grassroots systems provide a necessary complement: standardized evidence and second-review confirmations can feed into official workflows ({\Gg{7}}), while organizer–researcher collaboration may refine AI-style signatures ({\Gg{8}}). These lessons underpin our chess-inspired framework for sustainable co-governance. 

\subsubsection{Speculative Ideas and Future Directions}

Interviewees proposed both practical and risky ideas to improve fairness. One was to “control the environment.” {\Ppp{5}} suggested platform-supplied IDEs or hardened browsers so all contestants code in the same audited setup (pinned compilers, reproducible builds, basic telemetry). The expected gains were comparability and fewer covert channels for model use. Yet {\Gg{3}} argued that deploying and maintaining such clients globally would be costly, fragile across OS and localization, and likely to trigger an endless cheater–detector arms race. Both also noted equity risks (bandwidth, device performance, institutional firewalls) that could burden some contestants disproportionately.  

Another idea was to “watch the person, not just the code.” {\Tt{7}} proposed learning a per-contestant style profile over time and flagging sudden drifts in features such as identifiers, comment cadence, or structural idioms. She cautioned, however, that drift can be benign (e.g., language switches, editor changes, new templates), so alerts should be conservative, opt-in, and appealable, with clear documentation. Several interviewees emphasized low false-positive rates, calibrated thresholds, and transparent post-hoc explanations when flags are raised.  

A third line pointed to incentives. {\Cc{12}} argued that over-reliance on a single leaderboard in hiring fuels cheating markets. He called for broader signals—portfolios, supervised projects, multi-platform activity, peer attestations—so gaming ratings would no longer pay off. Others suggested de-emphasizing headline ratings in favor of richer longitudinal views (participation stability, editorial contributions, post-contest write-ups) to make reputation harder to manipulate.  

A compact compromise was to run two tracks—one AI-permitted, one human-only. The former would allow bounded, disclosed use; the latter retain hard bans with light post-hoc checks. But {\Gg{3}} warned that an AI track entrenches resource gaps (premium models, credits, low-latency compute), while {\Gg{5}} added that a human-only track still faces old challenges—undetected misuse, false positives, identity churn—so disclosure, separate ratings, and due-process appeals would be needed.  

Two pilot ideas also surfaced. {\Cc{3}} proposed an “audit lottery”: after each round, a random slice of accepted submissions would face a short oral/code check, anchored by optional locally hashed scratchpad notes. {\Gg{5}} suggested a lightweight community-evidence intake with reputation for accurate reporters and a time-boxed disclosure form for minor aids that do not replace algorithmic reasoning. Both proposals stressed bounded burden, due-process protections (ownership checks, appeal windows), and pre-registered evaluation metrics (precision/recall of flags, resolution latency, contestant trust) before any scale-up.

Seven interviewees drew explicit parallels between mind sports and today’s programming contests. As \T{3} put it: “At its core, competitive programming is a kind of chess—the computer is the board, problem setters lay out multiple games at once, and solvers win by producing the largest number of correct solutions in the shortest time.” \Pp{5} likewise emphasized the strong game-theoretic flavor of setter–solver interaction in programming contests.

\Pp{1} argued that the governance model used in chess communities—across both online and over-the-board play—maps well onto fairness oversight for programming contests: “I agree with Professor Saining Xie that the AlphaGo moment for programming hasn’t arrived yet~\cite{Xie2025AlphaGo}, but AI already defeated top human players back in 2016. In one sense Go is ‘solvable’ for AI, yet the sport did not disappear. Tournaments continue to run smoothly online and offline, and platform credibility has held up. I think there is a great deal our community can learn from that.”

Building on these reflections, several interviewees argued that programming contests could learn from chess governance. Chess federations survived the AI revolution by clarifying permitted tools, separating human-only and computer-assisted formats, and introducing random audits and disclosure. Rather than collapsing under AI dominance, the game stabilized through shared rules and visible enforcement. By analogy, contests might adopt clearer divisions between tracks, disclosure norms, and community evidence under due-process review—parallels that motivate our chess-inspired framework.  

\subsubsection*{Answer to RQ3}

Governance of AI in contests is already hybrid, co-produced by organizers and communities. Official measures supply structure and sanctions but face limits of scale, cost, and accuracy: our platform-level analysis shows that severe sanctions $R^{\mathrm{severe}}_c$ remain rare and false positives are vanishingly uncommon, yet behavioral proxies and grassroots investigations reveal that AI-assisted and scripted workflows extend far beyond what official logs capture. Grassroots initiatives add agility, probes, and second-review safeguards, but remain resource-constrained and sometimes contested. Iteration is ongoing: platforms refine bait prompts and heuristics, while volunteer moderators embed due-process protections and calibrate punishment norms. Across roles, interviewees agreed that perfect prevention is infeasible; credibility instead depends on transparent procedures, reversible decisions, and continual calibration in the face of shifting tools.

In sum, AI rules evolve through this hybrid ecosystem. Contestants broadly accept authority, but the compliance–trust gap shows that legitimacy rests on enforcement that is both visible and fair: systems must be strict enough to matter, yet accurate enough that being flagged still feels procedurally legitimate, as in the striking case of the interviewee who chose not to appeal despite being sanctioned. Inspired by chess—where clear tool rules, dual formats, and transparent adjudication have helped preserve credibility—we outline a governance framework emphasizing layered enforcement, auditable boundaries, and procedurally disciplined community participation as a path to safeguarding fairness and trust in the AI era.

\section{Discussion}

\subsection{Crossroads of Education, Competition, and Creation}

Programming-contest platforms sit at the crossroads of several AI-affected online communities. They function at once as learning environments, competitive arenas, and creative ecosystems around problem design and editorial writing. To reason about how they should respond to generative AI, it is useful to situate programming contests among three better-studied regimes: educational and assessment platforms, competitive online games and e-sports, and creative or fandom communities.

Prior work on educational and assessment platforms shows a strong emphasis on academic integrity and equal access: institutions experiment with hard bans in proctored exams, tightly scoped allowances in homework, disclosure requirements, and honor codes that try to preserve learning while recognizing that students will use AI tools anyway~\cite{10.1145/3613904.3642773,10.1145/3626252.3630958}. Some systems integrate plagiarism and AI-detection services into their workflows, yet recent studies report higher evasion rates and persistent concerns about detector accuracy, bias, and fairness~\cite{Novak2019Source-code,10.1145/3428206,gritsai2025are,11029804}.

In competitive online games and e-sports, platforms treat AI assistance and automation primarily as a threat to real-time competitive integrity. Responses include client-side anti-cheat software, server-side anomaly detection on input and timing patterns, escalating sanction ladders, and sometimes publicly visible ban waves to signal deterrence.~\cite{Cho2024Unpacking} Here, the focus is less on learning and more on protecting rankings, prize pools, and the legitimacy of spectator competition.~\cite{Mulakaluri2024Reviewing}

Creative and fandom communities take yet another stance. Rather than talking about “cheating,” they debate authorship, attribution, and the cultural legitimacy of AI-assisted works.~\cite{Li2024Fandom} Common strategies include separate tags or sections for AI-generated content, opt-in or mandatory disclosure of AI use, community guidelines for crediting data sources and models, and, in some cases, complete bans in subspaces that foreground human craft.~\cite{Formosa2024Can}

Our findings suggest that programming-contest platforms sit at the crossroads of all three regimes: they are at once learning environments, competitive arenas, and creative ecosystems built around problem design and editorial writing. This hybrid identity implies that no single imported governance model will suffice. Instead, contest communities may need to combine educational logics (protecting learning), competitive logics (safeguarding rankings), and creative logics (valuing disclosure and credit), and to be explicit about which facet they prioritize when making decisions about AI. Among these AI-affected communities, mind sports such as chess and Go are structurally closest to programming contests: they share Elo-style ratings, tournament formats, and a long history of negotiating computer assistance. We therefore next turn to chess and Go as concrete governance templates, asking what lessons their layered responses to AI can offer for programming contests.

\subsection{Chess-inspired governance approach}
Building on this cross-community view, mind sports offer a particularly close analogue to programming contests. Artificial intelligence has already surpassed human champions in many mind sports, from Deep Blue’s victory over Kasparov~\cite{campbell2002deep} to AlphaGo’s domination of Go~\cite{granter2017alphago}. Yet despite this asymmetry, chess and Go communities have preserved competitive legitimacy through layered governance~\cite{iliescu2020impact}, evolving technical detection systems, institutional policies, and community norms that sustain fairness~\cite{bart2021competitive,10.1007/978-3-031-34017-8_14}. This trajectory offers a useful analogy for programming contests, where LLMs now threaten ratings and credibility.

In chess, fairness rests on algorithmic detection and human judgment. Regan’s intrinsic performance rating compares moves against engine predictions and Elo, flagging anomalies for review~\cite{regan2011intrinsic}. Notably, the Elo system~\cite{pelanek2016applications}, also used by platforms like Codeforces, makes this logic transferable. Platforms complement such methods with ML classifiers on game features and time-use patterns, but crucially expert arbiters make final calls to mitigate false positives. Go federations similarly rely on meticulous game-by-game expert review, even at “99\% certainty,” to avoid punishing innocents~\cite{Bilen2020Online}.

Community-driven mechanisms are equally important. Research shows that empowering users to flag violations enhances legitimacy and distributes oversight~\cite{Zuckerman2023From}. Chess and Go operationalize this through reporting systems, volunteer moderators, and even third-party blacklists, while transparency reports reinforce norms that “fair play” is shared responsibility.

These lessons migrate readily to programming contests. ICPC and IOI already require finalists to declare online platform ratings, acknowledging continuity between online and offline performance. A chess-inspired governance approach would formalize this linkage: organizers could cross-reference trajectories with offline results, as FIDE compares intrinsic ratings to Elo. Independent cheat databases on Codeforces echo bottom-up accountability and could be integrated into official pipelines. Novel adaptations are possible too—VR monitoring proposed in Go~\cite{nordicgo2020vr} could enable distributed yet proctored contests.

Ultimately, online cheating cases in Go highlight that the migration of traditional mind sports online introduces unfairness that destabilizes rules. Solutions often borrow from digital-native esports, where cheats and countermeasures co-evolved, producing robust practices~\cite{hamlen2015problem,gabbiadini2014interactive} Programming contests, though born in computing, are only now facing their most formidable “cheatware”: LLMs. Balancing their disruptive force, and retracing the path chess and Go took against AI, remains a central challenge.

Drawing on lessons from chess and Go, we outline several directions for programming contests to preserve fairness in the era of LLMs:

\begin{itemize}
    \item \textbf{Rating-linked anomaly detection.} Contest platforms should implement cross-checking pipelines that compare online Elo-style ratings~\cite{pelanek2016applications} with offline ICPC/IOI outcomes. For example, a sudden 500+ rating jump online but consistently low offline performance could automatically trigger review, similar to Regan’s intrinsic rating method in chess~\cite{regan2011intrinsic}.
    \item \textbf{Expert review for borderline cases.} Platforms can establish independent review boards composed of experienced setters and coaches. Much like chess arbiters adjudicate disputed engine-detection cases~\cite{Bilen2020Online}, these boards would examine flagged contestants’ submission histories and training code to minimize false positives.
    \item \textbf{Community-driven oversight.} User flagging and third-party cheat databases (e.g., Codeforces initiatives) can be incorporated as auxiliary signals~\cite{Zuckerman2023From,Laarhoven2022Towards}. To ensure legitimacy, community reports should feed into, but not replace, formal review, creating a balance between grassroots accountability and institutional authority.
    \item \textbf{Proportional yet rigorous sanctions.} A transparent sanction ladder should be published—from temporary account suspension to permanent bans—with decisions accompanied by anonymized evidence summaries. This mirrors escalation practices in chess federations while reinforcing credibility and deterrence.
\end{itemize}

Together, these measures extend the layered governance of mind sports to programming contests, balancing automation, expertise, and community norms to safeguard credibility.

\subsection{Cheating for Credentials? Rethinking the Educational Value of Competitive Programming}

Incorporating competitiveness and playfulness into teaching has been shown to enhance instructional quality, encourage student exploration, and build confidence~\cite{Wang2023Exploring}. The original purpose of competitive programming contests was likewise to popularize computer science across countries, regions, women, and technologically disadvantaged areas~\cite{sun2013acm}. Today, however, competitive programming has also become a powerful mechanism for talent selection in both academia and industry. In our interviews, we learned that in some regions, a high Codeforces rating can even help undergraduate students secure well-paid jobs directly. However, such instrumentalization has also emerged as a key driver of cheating, which undermines the integrity of the contests themselves. \T{4} observed that more and more students join university competitive programming teams for utilitarian purposes, noting that while prizes and recognition may attract participants, educators, platforms, and contestants themselves should also recognize why earlier champions gained the respect of academia and industry. As \T{1} explained, “In industrial applications, the programming techniques from competitive programming are almost never directly useful. But we like hiring competitive programming contestants not because of their raw coding skills, but because of their passion for competitive programming, their teamwork, stress tolerance, and rigorous research attitude, which resonate with both academia and industry.” When increasing numbers of contestants exploit AI to obtain inflated scores that do not reflect their true abilities, they betray the very spirit of competitive programming. Educators must continually emphasize, as students engage deeply with the competitive programming community, that their participation should be grounded in genuine purpose, rather than allowing them to indulge in honors that are undeserved and obtained through deception~\cite{Miller2018The}.

\subsection{Beyond Leaderboards — AI Models and Their Role in STEM Competitions}

\textbf{Science, Technology, Engineering, and Mathematics (STEM)} contests such as \textbf{ICPC}, \textbf{IOI}, and \textbf{IMO (International Mathematical Olympiad)} were never intended as publicity platforms for corporate benchmarks. Their mission was to foster disciplined practice, perseverance, teamwork, and shared norms, while broadening access to computing and mathematics across regions and demographics~\cite{breiner2012stem,sun2013acm}. Excellence signals years of effort and community participation, but does not map directly to production-grade engineering or broad mathematical competence; competitive programming tasks often diverge from industrial problem structures and collaboration dynamics.

However, because achievements in \textbf{STEM} contests are easily quantifiable and often equated with being a “genius student,” many technology companies and research teams have seized upon contest scores as a blunt instrument to promote their models’ mathematical, programming, or physical reasoning ability. Yet such publicity frequently overlooks a critical distinction between (i) achieving high scores under narrow protocols and (ii) demonstrating broad, transferable capability. The former does not imply the latter, given validity gaps such as curated tasks, orchestration, and multiple-try inflation~\cite{bart2021competitive}. Excessive media amplification—even premature leaderboard announcements without official evaluation—has generated negative public discourse\cite{mashable2025imo} Olympiad organizers therefore stress “verification before amplification”~\cite{IMOOfficial2025}: AI results must first be checked against official rules—task access, timing, resources, grading—before amplification through media or marketing, ensuring claims rest on certified procedures rather than premature headlines.

Unlike the headline-grabbing narratives of AI clashing with chess or Go grandmasters, safeguarding participants is another pressing concern. Many IOI and IMO contestants are minors, for whom “AI beats students” stories can foster unhealthy comparison, misdirect media attention, and create pressures they never consented to~\cite{UNICEF2021AIChildren}. Coaches and stewards warned that sensational announcements erode trust in rankings and divert organizers from education to public-relations disputes.

AI models can participate in STEM contests as research experiments, but communication must prioritize education. Evaluations should disclose constraints and uncertainties, be released in research-marked tracks, and avoid sensationalizing minors. Media should refrain from comparative slogans and instead focus on inviting youth into STEM. Industry should respect contest processes, protect young competitors, and communicate responsibly to uphold fairness and educational value. Many technology companies exploit public misunderstandings to market AI by equating competitive programming with AI capabilities. In several languages, "competitive programming" and "algorithm contests" are often confused, inflating AI's role in solving AI algorithmic problems. However, competitive programming is a core computer science contest unrelated to AI algorithm. As \Pp{7}, an algorithm engineer, warns, “The hype around AI risks entering the uncanny valley\cite{mori2012uncanny}, where public belief in AI solving AI problems overshadows human development.” Beyond the leaderboards, the role of AI in STEM competitions should be about supporting, not replacing, human innovation and education.

\subsection{Limitation and Future Work}

Our study is not without limitations. First, although both the survey and interviews intentionally solicited international perspectives across roles and rating tiers, the sample remains skewed toward East Asian and male participants. This partly reflects the demographic reality of competitive programming~\cite{Yamaguchi2024The,Steegh2019Gender}, where these groups constitute a majority, but nonetheless limits the inclusiveness of our findings. Future work will aim to expand recruitment among women, underrepresented regions, and multilingual communities to capture a more balanced picture.  

Second, while we successfully interviewed administrators from most major international contest platforms, due to special circumstances we were unable to interview the stewards of \textbf{Codeforces}, the largest and most influential platform. Including perspectives from this platform would provide critical insights, and future research should prioritize engagement with its leadership.  

Third, social desirability bias~\cite{grimm2010social} poses a key limitation: interview and survey respondents may downplay or conceal actual instances of AI-assisted cheating to maintain a morally favorable self-presentation. Consequently, our current qualitative data primarily reflects formal contest settings and likely underestimates misconduct. Given that competitive programming communities are deeply embedded within rich online cultures, practices of knowledge sharing, norm enforcement, and AI tool usage are often negotiated in social media spaces such as forums, Discord servers, and X (formerly Twitter). Future work could therefore complement interviews with large-scale social media analysis—scraping official announcements, cheating reports, and discussion threads across platforms—to better capture emergent discourse and governance practices.

Fourth, our multilingual data collection and translation procedures introduce potential sources of bias. We restricted our own use of GPT-4o to translation-only prompts, reviewed translations against the original text, and resolved cross-regional terminology mismatches by anchoring discussions in canonical English terminology. Even so, subtle meaning shifts may have occurred, and we cannot fully exclude undisclosed LLMs use by participants when drafting responses. Our follow-up checks increase confidence that the accounts reflect participants’ own perspectives, but minor stylistic or interpretive artefacts may remain.

Fifth, our platform-level statistics and behavioral proxies are necessarily imperfect. The sanction metrics we derive from Codeforces logs capture only officially recorded violations and cannot see undetected misconduct. Likewise, our Python-share metric $\rho^{\text{py}}_{c}$ is a coarse, noisy signal: language choice is shaped by long-term shifts toward Python in programming education and contest communities, so increases in $\rho^{\text{py}}_{c}$ cannot be attributed solely to AI-assisted or scripted workflows. We therefore interpret these quantities as descriptive trends rather than causal evidence of AI usage, and rely more heavily on the temporal-uniformity metric and steward accounts as conservative lower bounds on automation. Future work with richer instrumentation or longitudinal baselines (e.g., pre-LLM years) is needed to more cleanly disentangle general ecosystem change from AI-specific effects. 

Finally, our study took place during a period of rapid technological change. As LLMs capabilities evolve, perceptions and practices observed in 2025 may shift, calling for longitudinal and comparative studies to track how governance adapts across contest ecosystems. Our own reliance on GPT-4o as a translation aid also reflects the tensions we document: even when constrained to “literal” translation and followed by human review, LLMs blur boundaries between infrastructure and authorship. We therefore treat them as part of the sociotechnical condition under study, rather than a neutral background tool.

\section{Conclusion}

This paper examined how large language models are reshaping competitive programming, long defined by rapid iteration, feedback, and integrity norms. Drawing on 37 interviews and a global survey of 207 contestants, we traced shifting workflows, fairness boundaries, and governance practices. Our findings show that while LLMs accelerate training and post-contest review, they also increase incentives for misuse in high-stakes settings where ratings serve as credentials.

We highlight three contributions: (i) empirical insights into evolving workflows across contestants, setters, coaches, and stewards; (ii) cross-role perspectives on fairness, showing how boundaries between support and cheating are negotiated; and (iii) a chess-inspired governance framework combining anomaly detection, expert review, community oversight, and proportional sanctions. Beyond contests, we caution against sensationalizing AI achievements in STEM competitions: leaderboard claims risk conflating narrow scores with broad capability and distracting from educational missions, especially when minors are involved. Ultimately, programming contests should continue not to crown machines, but to invite young people into computing, cultivating resilience, collaboration, and community—values that must be preserved through rigorous governance in the AI era.

\section*{Disclosure about Use of LLM}

As noted in Section~3.1.5, we used GPT-4o to assist with translating non-English responses (Russian, Ukrainian, Portuguese, Arabic, and Korean) into English. When invoking GPT-4o for this purpose, we constrained it with prompts of the form: ``Translate the following \textbf{[source language]} text into English. Do not summarize, rewrite, or expand the content beyond translation.'' This step was taken to ensure accurate comprehension of participants’ intended meanings while preventing the model from generating new content. All translations were subsequently reviewed by the research team before being incorporated into our analysis or quotations.

As noted in Section~4, we also used GPT-4o to assist with writing Python scripts. It was employed to clean raw data and to generate Jupyter Notebook code for data visualization. GPT-4o did not intervene in the selection of data analysis methods, the construction of our codebook, or the interpretation of results.

\begin{acks}
This work was supported in part by the National Key Research and Development Program of China under Grant 2024YFC3307602, the Guangdong Provincial Talent Program, Grant No.2023JC10X009 and the Red Bird MPhil Program at The Hong Kong University of Science and Technology (Guangzhou). 

We are grateful to the \textbf{CCF NOI} team led by \textbf{Zide Du} for their constructive feedback on this paper, and we thank them for over three decades of contributions to the popularization, development, and rule-making of competitive programming in China.

We thank all participating competitive programming contestants, problem setters, coaches, and platform stewards for their contributions to our study. We also thank the companies AtCoder, Luogu, CodeChef, QOJ, and BOJ for their responses and cooperation.  While our experiments and conclusions may not be perfect, we are committed to working with the community to uphold fairness, transparency, and educational value in the age of LLMs. May every submission in your future contests bring with it an \textbf{Accepted}!
\end{acks}

\bibliographystyle{ACM-Reference-Format}
\bibliography{bib/refs}

\newpage
\appendix

\clearpage
\onecolumn
\section*{Appendix}
\section{Additional Participant Tables}
\phantomsection
\label{sec:appendix-tables}

\textbf{Years} indicates how long an interviewee has served in that role; the same person may therefore have different values across tables.

\small
\renewcommand{\thetable}{A\arabic{table}}

\subsection{Problem Setters (P1--P7)}
\phantomsection
\label{subsec:appendix-problemsetters}
\vspace{-4pt}
\setcounter{table}{0}
\begin{table}[H]
  \centering
  \captionsetup{type=table}
  \caption{Overview of Problem Setters (P1--P7). ``Also C\#'' denotes appearance in the contestant table.}
  \label{tab:A1-problemsetters}
  \begin{tabular}{|c|p{0.19\textwidth}|c|p{0.52\textwidth}|c|}
    \hline
    \cellcolor{ProblemSetterBg}\textbf{Problem Setter} & \cellcolor{ProblemSetterBg}\textbf{Country or Region} & \cellcolor{ProblemSetterBg}\textbf{Years} & \cellcolor{ProblemSetterBg}\textbf{Notes} & \cellcolor{ProblemSetterBg}\textbf{Gender} \\ \hline
    \Pid{1} & China & 10 & IOI/ICPC World Champion; problem-setting expert & Male \\ \hline
    \Pid{2} & USA   & 4  & Also \C{5} & Male \\ \hline
    \Pid{3} & China & 3  & Also \C{3} & Male \\ \hline
    \Pid{4} & India & 1  & Also \C{12}; Newbie setter & Male \\ \hline
    \Pid{5} & China & 5  & High School Computer Science Teacher & Male \\ \hline
    \Pid{6} & China & 9  & Senior Researcher in Computer Science & Male \\ \hline
    \Pid{7} & China & 3  & Recommendation Algorithm Engineer (Also \T{7}) & Female \\ \hline
  \end{tabular}
\end{table}

\subsection{Coaches (T1--T7)}
\phantomsection
\label{subsec:appendix-coaches}
\vspace{-4pt}
\begin{table}[H]
  \centering
  \captionsetup{type=table}
  \caption{Overview of Coaches (T1--T7).}
  \label{tab:A2-coaches}
  \begin{tabular}{|c|p{0.17\textwidth}|c|p{0.26\textwidth}|p{0.28\textwidth}|c|}
    \hline
    \cellcolor{CoachBg}\textbf{Coach} & \cellcolor{CoachBg}\textbf{Country or Region} & \cellcolor{CoachBg}\textbf{Years} & \cellcolor{CoachBg}\textbf{Best Achievement} & \cellcolor{CoachBg}\textbf{Occupation} & \cellcolor{CoachBg}\textbf{Gender} \\ \hline
    \Tid{1} & China & 4  & Students' best performance undisclosed & Expert Software Engineer & Male \\ \hline
    \Tid{2} & USA (Brazil) & 2 & North America finals, Latin America finals & University Lecturer & Male \\ \hline
    \Tid{3} & China & 16 & World Finals & Professor & Male \\ \hline
    \Tid{4} & China & 10 & Asia-Pacific Prelims Gold Medal & Associate Professor & Female \\ \hline
    \Tid{5} & China & 1  & Novice Learner & Undergraduate Student Coach & Female \\ \hline
    \Tid{6} & China & 9  & National Silver Medal & High School Computer Science Teacher & Male \\ \hline
    \Tid{7} & China & 2  & Novice Learner & Senior Search \& Recommendation Algorithm Engineer (Also \Pp{7}) & Female \\ \hline
  \end{tabular}
\end{table}

\subsection{Platform Stewards (G1--G8)}
\phantomsection
\label{subsec:appendix-platform}
\vspace{-4pt}

\begingroup
\scriptsize
\setlength{\tabcolsep}{3.5pt}
\begin{table}[H]
  \centering
  \captionsetup{type=table}
  \caption{Overview of Platform Stewards (G1--G8).}
  \label{tab:A3-platform}
  \begin{tabular}{|c|p{0.18\textwidth}|p{0.12\textwidth}|p{0.54\textwidth}|c|}
    \hline
    \cellcolor{PlatformBg}\textbf{Platform Manager} & \cellcolor{PlatformBg}\textbf{Country or Region} & \cellcolor{PlatformBg}\textbf{Official Admin?} & \cellcolor{PlatformBg}\textbf{Notes} & \cellcolor{PlatformBg}\textbf{Gender} \\ \hline
    \Gid{1} & China & Yes & Official admin; rule setter & Male \\ \hline
    \Gid{2} & Japan & Yes & Official admin; rule setter & Male \\ \hline
    \Gid{3} & China & Yes & Official admin; rule setter & Male \\ \hline
    \Gid{4} & India   & Yes & Official admin; rule setter & Male \\ \hline
    \Gid{5} & South Korea & Yes & Official admin; rule setter & Male \\ \hline
    \Gid{6} & China & Yes & Official admin; rule setter & Male \\ \hline
    \Gid{7} & USA   & No  & Self-volunteered regulator; also \C{5} & Male \\ \hline
    \Gid{8} & UK    & No  & Self-volunteered regulator; also \C{4} & Male \\ \hline
  \end{tabular}
\end{table}
\endgroup

\section{Survey Questionnaire}

\renewcommand{\thetable}{B\arabic{table}}

\begingroup
\setcounter{table}{0}
\begin{table}[H]
\centering
\caption{Survey instrument -- demographics and background }
\label{tab:instr-demographics}

\begin{tabular}{|c|p{0.58\textwidth}|p{0.28\textwidth}|p{0.10\textwidth}|}
\hline
\rowcolor[HTML]{F2F2F2}
\textbf{ID} & \textbf{Item } & \textbf{Response type} & \textbf{Optional?} \\ \hline
1  & participant\_id & internal id (not shown to participants) & -- \\ \hline
\rowcolor[HTML]{FAFAFA}
2  & What is your current Codeforces rating? & numeric (self-reported rating) & No \\ \hline
3  & Which country or region do you reside in? & single choice (country/region) & No \\ \hline
\rowcolor[HTML]{FAFAFA}
4  & How many years have you been participating in competitive programming contests? & integer (years) & No \\ \hline
5 & Your Gender & single choice & No \\ \hline
\rowcolor[HTML]{FAFAFA}
6 & Do you believe the cf rating published by tech companies(eg. OpenAI) & Yes/No (model-rating trust) & No \\ \hline
7 & Your Email & short text (email) & \textbf{Yes} \\ \hline
\rowcolor[HTML]{FAFAFA}
8  & Have you ever used LLM tools (e.g., ChatGPT) during training or contests? & \textbf{Yes/No} & No \\ \hline
\end{tabular}
\end{table}

\begin{table}[H]
\centering
\caption{Survey instrument -- LLM usage and workflow items (Likert 1--5)}
\label{tab:instr-usage}
\begin{tabular}{|c|p{0.58\textwidth}|p{0.28\textwidth}|p{0.10\textwidth}|}
\hline
\rowcolor[HTML]{F2F2F2}
\textbf{ID} & \textbf{Item } & \textbf{Response type} & \textbf{Optional?} \\ \hline
9  & I frequently use LLM tools (such as ChatGPT) when practicing programming contest problems. & Likert 1--5 & No \\ \hline
\rowcolor[HTML]{FAFAFA}
10  & Integrating LLM assistance into my training has made me more efficient at solving problems. & Likert 1--5 & No \\ \hline
11  & LLM tools have become a regular part of my competitive programming workflow. & Likert 1--5 & No \\ \hline
\rowcolor[HTML]{FAFAFA}
12  & When I get stuck on a practice problem, I turn to an LLM for hints or solutions. & Likert 1--5 & No \\ \hline
13 & Using LLM assistance has improved my performance in programming contests. & Likert 1--5 & No \\ \hline
\rowcolor[HTML]{FAFAFA}
14 & LLMs help me learn new algorithms or programming techniques faster during my contest preparation. & Likert 1--5 & No \\ \hline
15 & Since I started using LLM tools, I have changed how I prepare for contests. & Likert 1--5 & No \\ \hline
\rowcolor[HTML]{FAFAFA}
16 & I am concerned that relying on LLMs too much might hinder the development of my own problem-solving skills. & Likert 1--5 & No \\ \hline
\end{tabular}
\end{table}

\begin{table}[H]
\centering
\caption{Survey instrument -- fairness, governance and policy items (Likert 1--5)}
\label{tab:instr-governance}
\begin{tabular}{|c|p{0.58\textwidth}|p{0.28\textwidth}|p{0.10\textwidth}|}
\hline
\rowcolor[HTML]{F2F2F2}
\textbf{ID} & \textbf{Item } & \textbf{Response type} & \textbf{Optional?} \\ \hline
17 & Allowing LLM use during contests gives certain participants an unfair advantage. & Likert 1--5 & No \\ \hline
\rowcolor[HTML]{FAFAFA}
18 & Using AI assistance during an official contest is essentially a form of cheating. & Likert 1--5 & No \\ \hline
19 & I am willing to comply with whatever rules contest organizers set about AI tool usage. & Likert 1--5 & No \\ \hline
\rowcolor[HTML]{FAFAFA}
20 & I trust that the current contest rules on AI/LLM usage are fair and appropriate. & Likert 1--5 & No \\ \hline
21 & All AI assistance should be completely banned during programming contests. & Likert 1--5 & No \\ \hline
\rowcolor[HTML]{FAFAFA}
22 & It's acceptable to allow only minor AI help, such as code autocompletion, but not full problem solutions from an AI. & Likert 1--5 & No \\ \hline
23 & Beginner-level contests should prohibit any AI tool use, while advanced-level contests could allow some AI assistance. & Likert 1--5 & No \\ \hline
\rowcolor[HTML]{FAFAFA}
24 & Contestants should be required to disclose after the contest if they used any AI assistance. & Likert 1--5 & No \\ \hline
25 & I am aware of the rules or policies that major contest platforms have regarding the use of AI tools in contests. & Likert 1--5 & No \\ \hline
\rowcolor[HTML]{FAFAFA}
26 & Contest platforms clearly communicate their policies on LLM/AI usage to participants. & Likert 1--5 & No \\ \hline
27 & The process by which contest platforms create or update rules on AI tool usage is transparent to participants. & Likert 1--5 & No \\ \hline
\rowcolor[HTML]{FAFAFA}
28 & I feel that participants have a say in how AI usage rules are set. & Likert 1--5 & No \\ \hline
29 & I have participated in community discussions about the use of AI tools in competitive programming. & Likert 1--5 & No \\ \hline
\rowcolor[HTML]{FAFAFA}
30 & I trust contest organizers to enforce AI-related rules fairly and effectively. & Likert 1--5 & No \\ \hline
31 & Overall, I am satisfied with how competitive programming platforms are handling the issue of AI tool usage. & Likert 1--5 & No \\ \hline
\end{tabular}
\end{table}

\begin{table}[H]
\centering
\caption{Survey instrument -- open-ended prompts (optional)}
\label{tab:instr-open}
\begin{tabular}{|c|p{0.58\textwidth}|p{0.28\textwidth}|p{0.10\textwidth}|}
\hline
\rowcolor[HTML]{F2F2F2}
\textbf{ID} & \textbf{Item } & \textbf{Response type} & \textbf{Optional?} \\ \hline
32 & In addition to the scenarios mentioned above, please describe any other ways you have used LLMs (AI tools) during your training or in contests. & free text (open-ended) & \textbf{Yes} \\ \hline
\rowcolor[HTML]{FAFAFA}
33 & Do you have any suggestions or concerns regarding how contest platforms should manage or enforce rules about AI (LLM) usage? & free text (open-ended) & \textbf{Yes} \\ \hline
\end{tabular}
\end{table}

\par\small\textit{Note: all ``Likert 1--5'' items use a 5-point agreement scale (1=Strongly disagree, 5=Strongly agree).}

\endgroup

\section{Formal Definitions of Platform-Level Metrics}
\phantomsection
\label{sec:appendix-metrics}

\small

\noindent We formalize the Codeforces metrics used in Section~\ref{sec:platform-stats}.

\subsection*{C.1 Official sanction metrics}

For each contest $c$ with $N_c$ participants, the \emph{severe-cheater rate} $R^{\mathrm{severe}}_c$ is defined as
\setcounter{equation}{0}
\begin{equation}
    R^{\mathrm{severe}}_c
    =
    \frac{\bigl\lvert \{\,u \in c \mid \texttt{participantType}(u) = \text{OUT}{\_}\text{OF}{\_}\text{COMPETITION} \land \exists\, s \in u:\, \texttt{verdict}(s) = \text{SKIPPED} \,\} \bigr\rvert}{N_c},
    \label{eq:severe-rate}
\end{equation}
i.e., the proportion of participants removed from the standings because at least one of their submissions is marked \texttt{SKIPPED}.

The \emph{mild-cheater rate} $R^{\mathrm{mild}}_c$ is defined as
\begin{equation}
    R^{\mathrm{mild}}_c
    =
    \frac{\bigl\lvert \{\,u \in c \mid \texttt{participantType}(u) = \text{CONTESTANT} \land \exists\, s \in u:\, \texttt{verdict}(s) = \text{SKIPPED} \,\} \bigr\rvert}{N_c},
    \label{eq:mild-rate}
\end{equation}
capturing contestants who remain in the standings but have specific submissions invalidated.

\subsection*{C.2 Behavioral proxies for automation and AI assistance}

For each contest $c$, we define the \emph{Python share} $\rho^{\mathrm{py}}_c$ as
\begin{equation}
    \rho^{\mathrm{py}}_c
    =
    \frac{N^{\mathrm{AC,py}}_c}{N^{\mathrm{AC,total}}_c},
    \label{eq:py-share}
\end{equation}
where $N^{\mathrm{AC,py}}_c$ is the number of accepted Python/PyPy submissions and
$N^{\mathrm{AC,total}}_c$ is the total number of accepted submissions.

For each contestant $u$ in contest $c$ with $k \ge 3$ accepted submissions at times
$T_{u,c} = \{ t_1 < \dots < t_k \}$, let the sequence of time gaps be
$\Delta T_{u,c} = \{ \delta_i = t_{i+1} - t_i \}$. The coefficient of variation (CV) is
\begin{equation}
    CV(u,c)
    =
    \frac{\sigma(\Delta T_{u,c})}{\mu(\Delta T_{u,c})},
    \label{eq:cv}
\end{equation}
where $\mu$ and $\sigma$ denote the mean and standard deviation of the gap sequence. In the main text, we treat contestants with $CV(u,c) < 0.2$ as \emph{highly uniform} and those with $CV(u,c) < 0.05$ as \emph{extremely uniform} and aggregate these counts per contest and division as proxies for scripted or tool-driven workflows.

\end{document}